\documentclass[useAMS,usenatbib]{mn2e}
\usepackage{natbib}
\usepackage{txfonts}
\usepackage{multirow}
\usepackage{graphicx}

\usepackage{xcolor}
\usepackage{epsfig,graphics}
\usepackage{rotating} 
\usepackage{savefnmark,multirow,topcapt}

\newcommand {\be}{\begin{equation}}
\newcommand {\ee}{\end{equation}}  
\renewcommand{\theenumi}{\arabic{enumi}.}

\renewcommand{\deg}{^{\circ}}

\newcommand{\abs}[1]{
\left|{#1}\right|
}

\newcommand{\fu}{ 4U 1630-47}

\title[Multicolour baryonic jet model]{Thermal X-ray emission from a baryonic jet:\\a self-consistent multicolour spectral model }   
\author[Khabibullin, Medvedev \& Sazonov ]{
\smallskip
I. Khabibullin$^{1,2}$\thanks{khabibullin@iki.rssi.ru},
P. Medvedev$^{1}$\thanks{tomedvedev@iki.rssi.ru} and
S. Sazonov$^{1,3}$\thanks{sazonov@iki.rssi.ru} \\
$^{1}$Space Research Institute, Russian Academy of Sciences,
Profsoyuznaya 84/32, 117997 Moscow, Russia\\
$^{2}$Max-Planck-Institut f\"ur Astrophysik,
Karl-Schwarzschild-Str. 1, 85740 Garching bei M\"unchen, Germany\\ 
$^{3}$Moscow Institute of Physics and Technology, Institutsky
per. 9, 141700 Dolgoprudny, Russia}
\date{Received \today}
\begin{document}
\maketitle
\begin{abstract}

We present a publicly-available spectral model for thermal X-ray
emission from a baryonic jet in an X-ray binary system, inspired by
the microquasar SS 433. The jet is assumed to be strongly
collimated (half-opening angle $\Theta\sim 1\deg$) and
mildly relativistic (bulk velocity $\beta=V_{b}/c\sim0.03-0.3$). Its
X-ray spectrum is found by integrating over thin slices of constant
temperature, radiating in optically thin coronal
regime. The temperature profile along the jet and corresponding
differential emission measure distribution are calculated with full
account for gas cooling due to expansion and radiative losses. Since
the model predicts both the spectral shape and luminosity of the jet's
emission, its normalisation is not a free parameter if the source
distance is known.

We also explore the possibility of using simple X-ray
observables (such as flux ratios in different energy bands) to
constrain physical parameters of the jet (e.g. gas temperature and
density at its base) without broad-band fitting of high-resolution
spectra. We demonstrate this approach in application to $Chandra$
HETGS spectra of SS 433 in its 'edge-on' precession phase,  
when the contribution from non-jet spectral components is expected to
be low. Our model provides a reasonable fit to the 1-3 keV
data, while some residuals remain at higher energies, which may be
partially attributed to a putative reflection component. 

Besides SS 433, the model might be used for describing jet components
in spectra of other Galactic XRBs (e.g. 4U 1630-47), ULXs (e.g. Holmberg II X-1), and candidate SS 433 analogues like S26 in NGC7793 and the radio transient in M82.

\end{abstract}
\section{Introduction}
\label{s:intro}

~~~~~ Launching of relativistic outflows from inner regions of an accretion disc is one the basic predictions of the standard theory of disc accretion for specific regimes when radiation alone is not efficient enough to carry away all the liberated potential energy of the in-falling matter \citep{SS1973}. Although there are numerous indications of such outflows both in X-ray binary systems (XRB) and active galactic nuclei (AGN), their actual composition (leptonic or baryonic) as well as the corresponding launching mechanisms (magnetic field vs. radiation pressure driven) still remain unclear (\cite{Fender2006}; see also \cite{Fender2014} and \cite{Worrall2009} for recent reviews in the context of XRBs and AGN,  respectively). While leptonic models are typically successful in describing the observed emission as having a primarily synchrotron origin, the existence of some baryonic loading is also frequently invoked to explain the high jet kinetic power required to produce and maintain large-scale cavities or cocoons associated with XRBs and AGNs \citep{Gallo2005,Dunn2005}.

The main evidence for the existence of baryonic jets comes, however, from the outstanding Galactic X-ray binary system SS 433, whose X-ray emission is marked by pairs of oppositely Doppler-shifted ($ V\sim 0.1 c$) lines of highly ionised heavy elements on top of the thermal bremsstrahlung continuum with $ T\sim 20 $ keV \citep{Kotani1996}, while its optical emission also bears similar signatures of Doppler-shifted lines of hydrogen and helium (e.g. \cite{Borisov1987}). These peculiar properties are attributed to a pair of strongly collimated (the half-opening angle $ \Theta\sim 1\deg$), mildly-relativistic (the bulk velocity $ \beta\approx0.26 $) precessing jets of ordinary highly ionised plasma with almost solar abundance of elements (see \citealt{Fabrika2004} for a review). According to the standard multi-temperature model, the X-ray jets first become visible when their temperature is $ T_0\sim 20 $ keV  and then cool below $ T\sim 1$ keV as a result of adiabatic expansion, which allows a number of ionization stages of various elements to contribute to the net observed spectrum \citep{Kotani1996}. 

By now, SS 433 remains as prominent as unique, with no other source known to unequivocally demonstrate the presence of baryonic jets. Recently, \cite{Diaz2013} claimed the detection of a pair of Doppler-shifted lines of H-like iron in the spectrum of \fu. These lines, however, were not detected in subsequent observations by \cite{Neilsen2014}. Another promising opportunity comes from observations of ultra-luminous X-ray sources (ULXs), at least some of which are believed to be SS 433-like systems, observed nearly 'face-on' \citep{Fabrika2001,Begelman2006} (see e.g. an analysis of Holmberg II X-1 spectrum by \cite{Walton2015}). To look for the presence of baryonic jets in such cases, it is desirable to have a spectral model that would consistently predict the emergent X-ray spectrum for a broad range of jet parameters, and it is natural to build such a model based on the properties of the prototypical jets in SS 433. 

Plenty of high quality X-ray spectroscopic data is now available for SS 433 from the \textit{Chandra} \citep{Marshall2002,Namiki2003,Lopez2006,Marshall2013}, \textit{XMM-Newton} \citep{Brinkmann2005,Medvedev2010} and \textit{Suzaku} \citep{Kubota2010} observatories at various phases of precessional and orbital motion. However, the standard jet model usually fails to reproduce the observed complexity of the spectra (e.g. \cite{Brinkmann2005}). Different ways have been proposed to eliminate the discrepancies, based on either adding some additional components \citep{Medvedev2010} or further elaboration of the standard model with the account for radiative cooling, radiation transfer effects and deviations from the coronal approximation (\cite{KS2012}). The level of sophistication of these models inevitably causes the loss of generality, making them hardly applicable to any other source that can potentially possess baryonic jets (like Galactic XRB \fu \citep{Diaz2013}, ULX Holmberg II X-1 \citep{Walton2015}, or candidate SS 433-analogues like S26 in NGC7793 \citep{Soria2010} and the unusual radio transient in M82 \citep{Joseph2011}).   

Here we present a model\footnote{The model with some related material is available for downloading at the web-site ~\texttt{http://hea133.iki.rssi.ru/public/bjet/}
.} intended to serve as a missing link between the well-studied case of SS 433 and sources with suspected presence of baryonic jets. A compromise between generality and sophistication is achieved by i) considering a rather broad range of jet parameters and ii) modifying the standard multi-temperature jet model by self-consistently treating radiative cooling and taking into account flux redistribution in triplets of He-like ions due to collisional de-excitation (while no account of radiative transfer effects and other possible deviations from the coronal approximation (see \cite{KS2012,Marshall2013}) is taken).    

Section \ref{s:model} describes this model in detail, while in Section \ref{s:observ} we discuss basic properties of the predicted spectra and their sensitivity to jet parameters. In Section \ref{s:ss433}, the ability of the model to fit high-resolution spectroscopic data of SS 433 (provided by $Chandra$ HETGS) is verified. We end up with drawing conclusions in Section \ref{s:conclusions}. 
\section{Model}
\label{s:model}

A multicolour jet model with account for adiabatic expansion and radiative cooling was constructed numerically by \cite{Brinkmann1988} and analytically by \cite{Koval1989} (who did not include line emission either  in the calculated spectrum or in the radiative cooling term). Most recently, \cite{Medvedev2010} and \cite{KS2012} performed self-consistent calculations including both line and continuum emission based on up-to-date atomic data from the \texttt{AtomDB/APEC} model \citep{Foster2012}. Here, we follow the scheme and main designations adopted in \cite{KS2012}.
\subsection{Physical model}
\label{s:physmodel}
\begin{figure}
\centering
\includegraphics[]{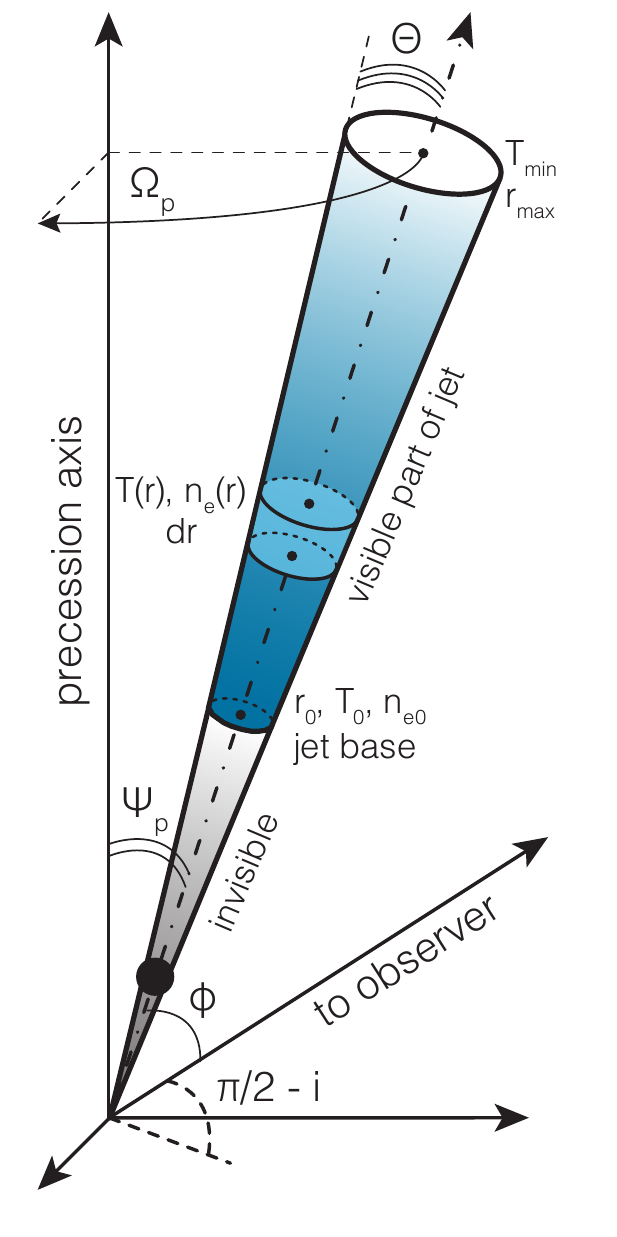}
\caption{Schematic view of the assumed jet's geometry. The jet is treated as an axis-symmetrical ballistic flow directed away from the compact object. The location of the latter is marked with a black dot and it may not coincide with the flow cone's apex. The half-opening angle of the flow cone is assumed to be rather small, $ \Theta \sim 1^{\circ}$, so the bulk velocity of the matter $v_b=\beta c$ is constant across the entire jet. The jet axis is inclined at angle $ \phi $ with respect to the line of sight, and $ \phi $ varies in time due to precession with amplitude $ \psi_p$ around an axis inclined at angle $ i$ to the line of sight. The precession period $ P_p$ is assumed to be much longer than the characteristic dynamical time-scale $ t_{d}\sim r_{0}/\beta~c $.
At the point $ r=r_0 $ where the jet first becomes visible to an observer, the gas temperature is designated as $ T_0 $ and electron number density as $ n_{e0}$. The end point $r= r_{max} $ corresponds to the position where gas temperature falls below $ \sim 0.1 $ keV (see text). }
\label{f:sketch}
\end{figure}

The X-ray jet is treated as an axis-symmetrical ballistic plasma flow
directed away from the compact object (see Fig.{\ref{f:sketch}), so
  the bulk velocity $ \beta$  of the matter is constant across the
  whole jet. For the problem in hand, we consider only
  mildly-relativistic jets, i.e. with  $ \beta\sim 0.1$ (see also
  Section \ref{ss:param}). Its degree of collimation is described by a
  half-opening angle $ \Theta $, which is assumed to be rather small,
  $ \Theta \sim 1^{\circ}\sim 0.02 $ radian, i.e. the jet is highly
  collimated. In this picture, the physical conditions inside the jet
  depend solely on the distance along its axis. Suppose that the jet
  first becomes visible to an observer at distance $ r_0 $ from the
  ballistic cone apex. This position is called the jet's base, and it
  is the place where initial conditions for the flow are to be
  set. Given electron number density $ n_{e0}$ and temperature $T_{0}
  $\footnote{ {The gas temperature $T$ is given in energy units
      (i.e. it actually equals $ \kappa_{B} T $) throughout the paper.}} there, the density and temperature profiles along the jet can be easily calculated. 

Indeed, the continuity equation implies constancy of the mass flow through the jet, i.e. 
\begin{equation}
\label{eq:mdot}
\dot{M}_{j}(r)=\dot{M}_{j}(r_0)=\mu m_{p}~(1+X)n_{e0}~\pi~ r_0^2\Theta^2 ~\beta c
\end{equation}
where $ \mu\approx 0.62$ is the mean molecular weight and $ X=n_i/n_e\approx 0.91$ is the ion-to-electron ratio. The cited values are appropriate for approximately solar chemical composition, but their actual sensitivity to the abundance of elements heavier than helium (hereafter metallicity) is rather weak. The metallicity is quantified by a factor $Z  $ relative to the set of solar abundances of \cite{Anders1989}\footnote{This abundance set is chosen simply because it is default for the \texttt{AtomDB} database, used in our calculations.}.

For fixed $ \Theta $ and $ \beta $, Eq.(\ref{eq:mdot}) implies $ n_{e}(r)~{r^2}=n_{e0}{{r_0}^2} $, so that
\begin{equation}
\label{eq:nprof}
n_{e}(r) = n_{e0}~\left(\frac{r}{r_0}\right)^{-2}.
\end{equation}
In order to find the temperature profile, one needs to solve the thermal balance equation :
\begin{equation}
\label{eq:cooling}
   \frac{dT}{dr}=-2\left(\gamma-1 \right)  \frac{T}{r}-\frac{2n_{e} n_{i}}{3\left(n_{e}+n_{i}\right)} \frac{\Lambda_{Z} \left( T \right)}{\beta c},
\end{equation}
where the first and second terms on the right-hand side correspond to  cooling due to adiabatic expansion and radiative cooling, respectively. The integrated emissivity $\Lambda_{Z} \left(T\right)=\int\epsilon_Z(E,T)dE$ is calculated in the coronal approximation for a hot optically thin plasma  (\texttt{AtomDB/APEC}, version 3.0.2, \cite{Foster2012}) with the specific value of $ Z $.
 {Note that Eq. (\ref{eq:cooling}) assumes that
  the temperatures of electrons and ions are in equilibrium, which can be justified by comparing the corresponding equilibration time-scale $ t_{eq}\sim 0.1~s~\times ~ T_{10}^{3/2}/n_{14} $ \citep{Spitzer1962} with the adiabatic $ t_{ac}\sim r_{0}/\beta c\sim 3~s~\times r_{11}/\beta $ or radiative (dominated by bremsstrahlung emission) $t_{cr} \sim T/n\Lambda(T)\sim 10~s~\times~T_{10}^{1/2}/n_{14}$ cooling time-scales, $T_{10}=T/10$ keV, $n_{14}=n_e/10^{14}$ cm$^{-3}$, $ r_{11}=r_0/10^{11}$ cm.}

By introducing dimensionless quantities $\eta$=T/T$_{0}$, $\xi$=r/r$_{0}$ and $\lambda_{Z,T_0}\left(\eta\right)$=$\Lambda_{Z} \left(\eta T_{0}\right)/\Lambda_{Z} \left(T_{0}\right)$
and assuming that $\gamma$=5/3, one obtains
\begin{equation}
\label{eq:unitlesscooling}
\frac{d\eta}{d\xi}=-\frac{4}{3}\frac{\eta}{\xi}-\alpha \frac{\lambda_{Z,T_0} \left(\eta \right)}{\xi ^{2}},
\end{equation}	
so that the temperature profile is governed by the parameter
\begin{equation}
\label{eq:alpha}
\alpha =  \frac{2}{3} \frac{n_{e0}r_{0}}{\beta c}\frac{\Lambda_{Z} \left(T_{0}\right)}{T_{0}} \frac{X}{1+X}.
\end{equation}

\begin{figure}
\centering
\includegraphics[width=1.1\columnwidth]{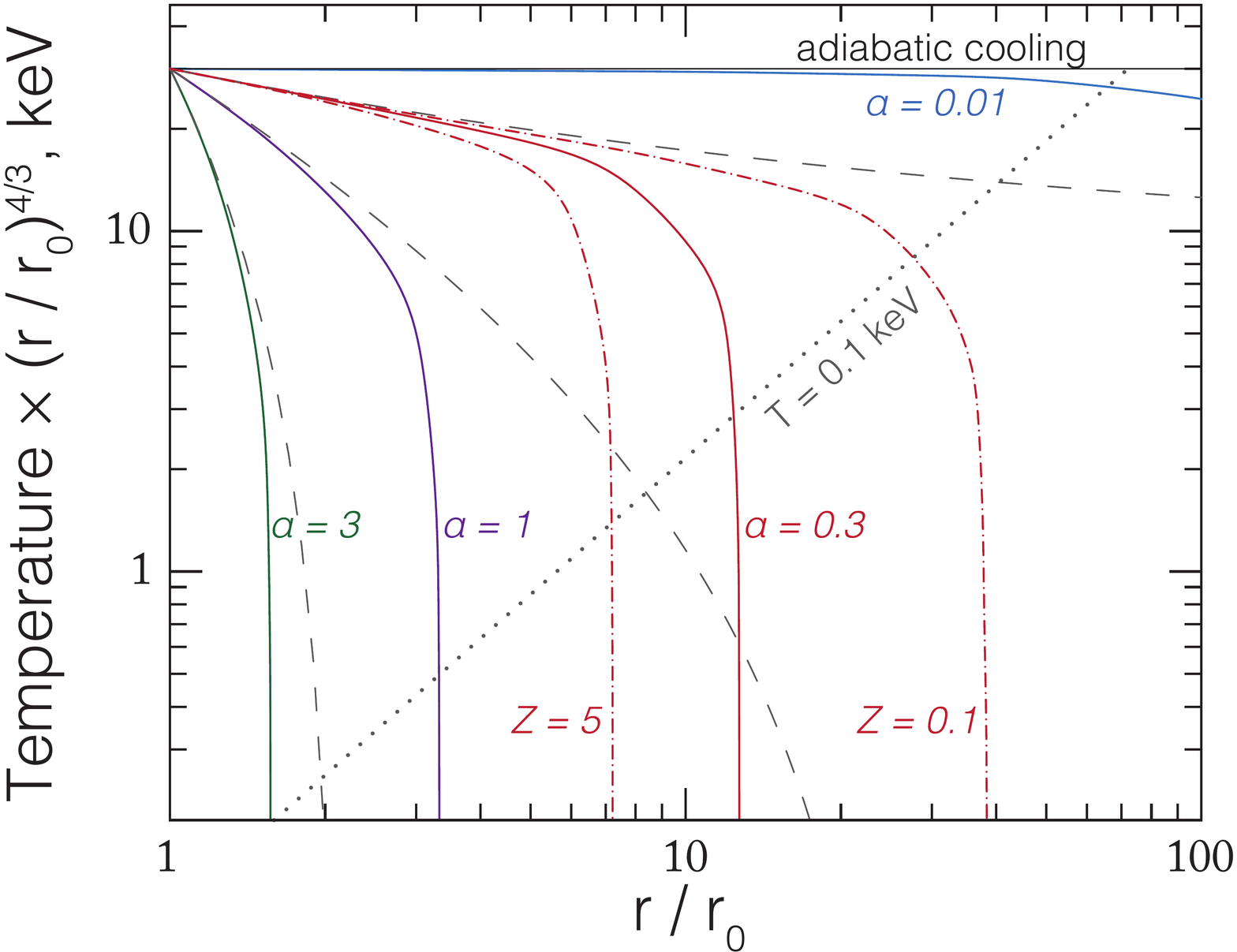}
\caption{Gas temperature profiles along the jet with $ T_0=25 $ keV multiplied by $ (r/r_0)^{4/3} $ for comparison with the adiabatic one. The solid curves correspond to the designated values of the parameter $ \alpha $ and $ Z=1$. The red dash-dotted lines show temperature profiles for $ \alpha=0.3 $ and $ Z=0.1$ and $ Z=5$. Black dashed curves demonstrate the profiles given by Eq.(\ref{eq:bremssol}), i.e. when only bremsstrahlung emission contributes to the radiative cooling term. The black dotted line marks the adopted jet boundary, i.e $ T=0.1$ keV.}
\label{f:tprof}
\end{figure}

Solving Eq.(\ref{eq:unitlesscooling}) numerically with the initial condition $ \eta\mid_{\xi=1}=1$ allows the temperature profile $\eta_{\alpha}\left(\xi\right)$ (or $T \left( r\right)=T_{0}\eta_{\alpha}\left(r/r_{0}\right)$) to be obtained. Since thermal instability is expected to evolve in the jet at $T\sim T_{min}\simeq 0.1 $ keV \citep{Brinkmann1988},  we consider only the hotter jet portion, i.e. $r<r_{max}=r_0 \xi_{\alpha}\left(T_{min}/T_0\right) $, where $\xi_{\alpha}\left(\eta\right)$ is the inverse function for $\eta_{\alpha}\left(\xi\right)$, which exists due to monotonicity of the latter. Temperature profiles for various values of $ \alpha$ are shown in Fig.\ref{f:tprof}.

As is clear from Eq.(\ref{eq:cooling}) and Eq.(\ref{eq:unitlesscooling}), the physical meaning of the parameter $ \alpha$ is essentially the ratio of the radiative cooling term to the adiabatic one  {at the jet's base}.  {Variation of this ratio along the jet is given by $ \mathcal{R}(\xi)=\frac{3}{4}\alpha ~ \frac{\lambda_{Z,T_0}(\eta)}{\eta~\xi} $ and is demonstrated in Fig.\ref{f:rprof}. For $\alpha \ll 1$, this ratio remains small across almost the whole jet, so the temperature profile must approach the adiabatic asymptote $ T\propto r^{-4/3} $ in this case}. 

Since radiative losses are dominated by bremsstrahlung emission for the high-temperature portion of the jet ($ T\gtrsim 5 $ keV for solar abundance of heavy elements), one may try to get an approximate solution assuming $ \Lambda_{Z}(T)\propto \sqrt{T}$, i.e. $ \lambda_{Z,T_0}(\eta)=\sqrt{\eta}$  in this region. Indeed, this assumption allows an exact analytical solution of Eq.(\ref{eq:unitlesscooling}) to be obtained \footnote{After substitutions $ u=2\sqrt{\eta}$ and $ \zeta=\ln\xi $, Eq.(\ref{eq:unitlesscooling}) reduces to $ \frac{du}{d\zeta}+\frac{2}{3}u=-\alpha e^{-\zeta} $, which is readily integrated 
.} (see also \cite{Koval1989}):
\begin{equation}
\label{eq:bremssol}
\sqrt{\eta}=\left[1-\frac{3}{2}\alpha\left(1-\xi^{-1/3}\right)\right]\xi^{-2/3},
\end{equation}
which, for $\alpha \gg 1$, is approximated by $\eta(\xi)=1-(\alpha+\frac{4}{3})(\xi-1)$ when $ \xi \rightarrow 1 $, and $ \eta=\frac{\alpha^2}{4}\left(1-\xi/\xi_{max}\right)^2 $ when  $ \xi \rightarrow \xi_{max}=\left(1-\frac{2}{3\alpha}\right)^{-3} $.
%
\begin{figure}
\centering
\includegraphics[width=1.1\columnwidth]{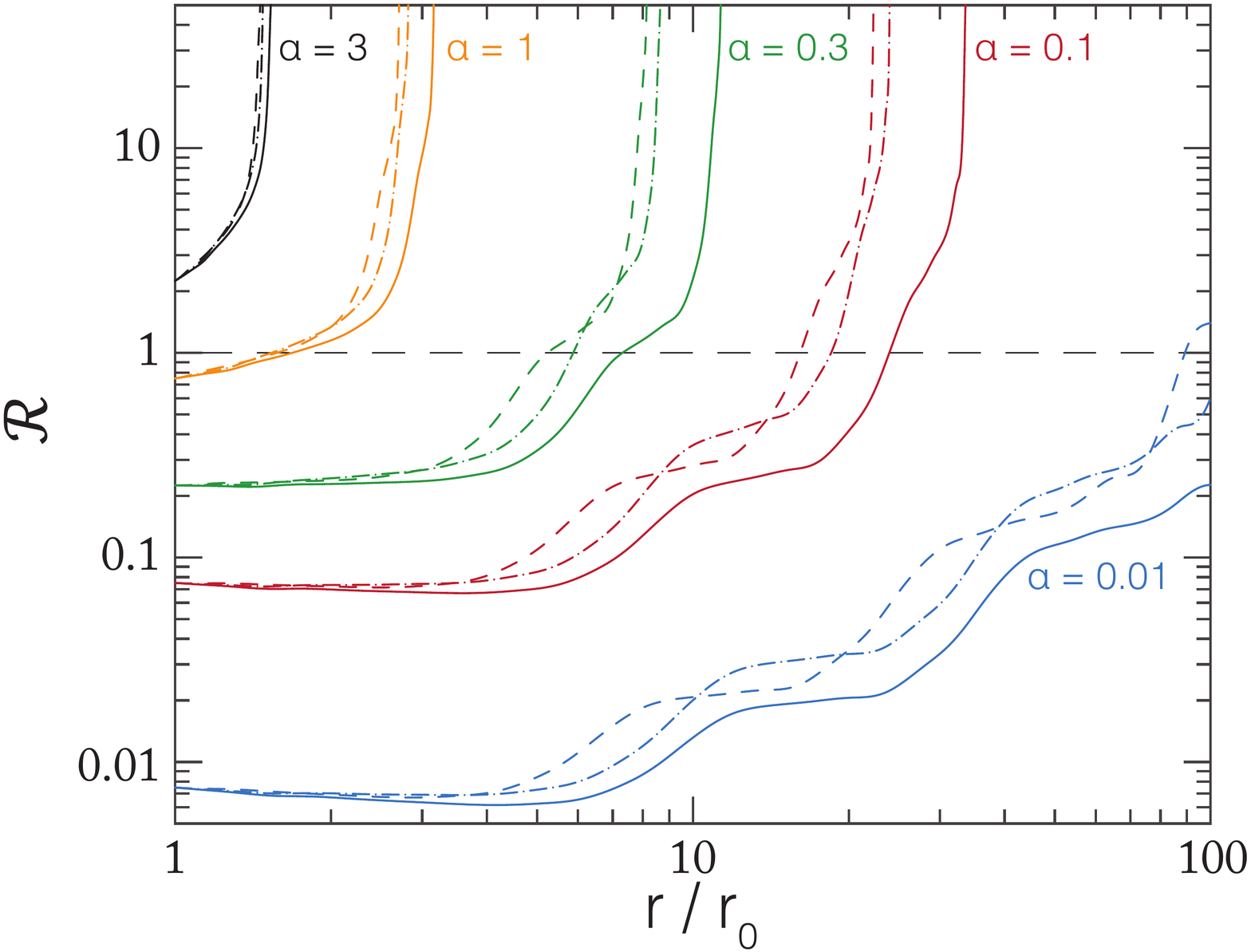}
\caption{ {Variation along the jet of the ratio of the radiative
    and adiabatic cooling terms for five different values of the parameter $ \alpha $ (as indicated on the plot). The solid curves correspond to $ T_0=25 $ keV and $ Z=1$, dashed curves to $ T_0=15 $ keV and $ Z=1$, and dash-dotted curves to $ T_0=25 $ keV and $ Z=2$. The horizontal long-dashed line marks the $\mathcal{R}=1$ level.}}
\label{f:rprof}
\end{figure}

As is shown in the next section, the absolute value of the logarithmic derivative $ \abs{\frac{d\ln \eta}{d\ln\xi}}$ is useful for calculation of the DEM distribution. For $ \alpha \ll 1 $, the temperature profile has a power-law shape, so that $\abs{ \frac{d\ln \eta}{d\ln\xi} }$ is constant and equals 4/3. For $ \alpha \gg 1 $, $\abs{ \frac{d\ln \eta}{d\ln\xi}}=\frac{\alpha}{\xi\sqrt{\eta}} $, as follows directly from Eq.(\ref{eq:unitlesscooling}) with $ \lambda_{Z,T_0}(\eta)=\sqrt{\eta} $.


\subsection{Spectrum calculation}
\label{ss:speccalc}
In the optically thin regime, one may calculate the emergent spectrum using the differential emission measure distribution, which is derived from the density and temperature profiles as
\begin{equation}
\label{eq:dem}
DEM(T)=\frac{dEM}{d\ln T}=\frac{n_e n_i~dV}{d\ln T}= X n_e^2\left(T\right)~\pi\Theta^2 r^2(T)\abs{\frac{dr}{d\ln T}},
\end{equation}
which is equivalent to 
\begin{equation}
\label{eq:demunitless}
DEM(\eta)=\pi X n_{e0}^2 r_{0}^3\Theta^2~\frac{1}{\xi(\eta)} \abs{\frac{d\ln\xi}{d\ln\eta}}.
\end{equation}
In view of the results of the previous section, one immediately gets 
\begin{equation}
\label{eq:demad}
 DEM(\eta)=\frac{3\pi}{4}~X n_{e0}^2 r_{0}^3\Theta^2~\eta^{3/4} \propto \eta^{3/4}
\end{equation}
for an adiabatic jet, and
\begin{equation}
\label{eq:demrad}
 DEM(\eta)= \frac{\pi}{\alpha}~X n_{e0}^2 r_{0}^3\Theta^2~\eta^{1/2} \propto \eta^{1/2}
\end{equation}
for the $ \alpha \gg 1 $ case  {with bremsstrahlung-dominated energy losses. The characteristic shapes of the DEM distribution are presented in Fig.\ref{f:demprof}.}

 {The result for $ \alpha \gg 1 $ may seem somewhat counter-intuitive recalling the very different shapes of the temperature profiles as a function of $ \alpha$ (see e.g. Fig.\ref{f:tprof}). In fact, it is a consequence of the jet being "short" in the $ \alpha \gg 1 $ limit, i.e. $ \xi_{max}-1 \ll 1$. Indeed, $ n_e $ is almost constant here, and $ dV(T)\propto dr(T)\propto t_{cr}(T) $, where $ t_{cr}(T) $ is the cooling time due to radiative losses. If losses due to bremsstrahlung emission dominate, $ t_{cr}(T) \propto \sqrt{T}$, and $ n^2dV(T) \propto \sqrt{T} $, so $  DEM(\eta)\propto \sqrt{\eta}$.}

\begin{figure}
\centering
\includegraphics[width=1.1\columnwidth]{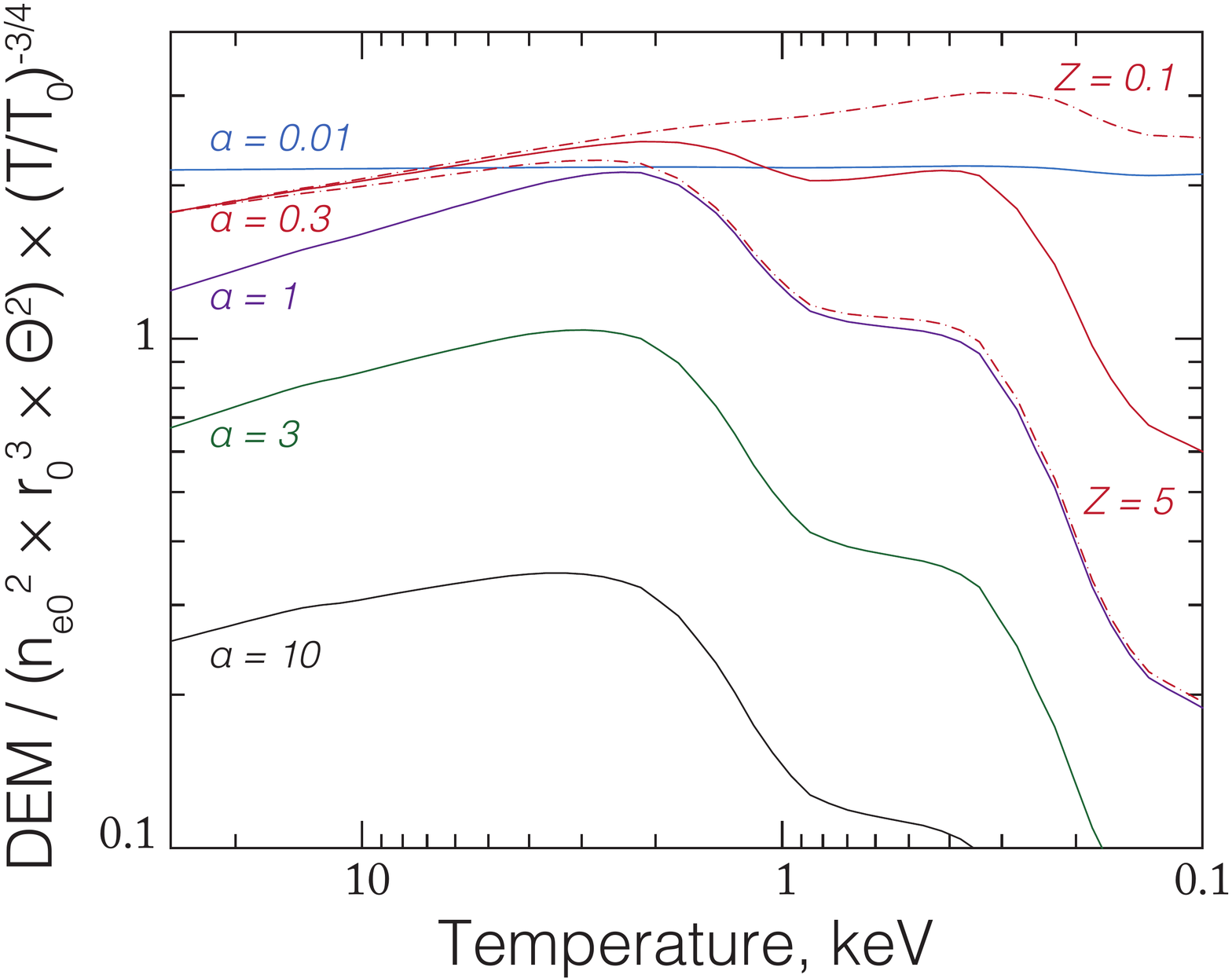}
\caption{Reduced differential emission measure (DEM) distributions for various values of the jet parameters. The original DEM has been divided by  $ n_{e0}^2 r_{0}^3 \Theta^3 $ in order to eliminate the normalisation dependency (see Eq.(\ref{eq:demunitless})) and multiplied by $ (T/T_0)^{-3/4} $ for comparison with the purely adiabatic case. At low $ \alpha $, the reduced DEM is consistent with the adiabatic one and does not  depend on $ \alpha $ (see Eq.(\ref{eq:demad})). At high $ \alpha $, the shape of the reduced DEM is determined by bremsstrahlung losses for $ T\gtrsim $ 3 keV and also has a power-law dependence there. For lower temperatures, cooling by line emission dominates, which is evident from the dependence of the DEM shape on the gas metallicity in this region (dashed-dotted curves show the $ Z=0.1 $ and $Z=5 $ cases for $ \alpha=0.3 $). The normalisation of the reduced DEM scales as $ \propto 1/\alpha $ for the high $ \alpha $ regime (see Eq.(\ref{eq:demrad})).}           
\label{f:demprof}
\end{figure}

In the comoving reference frame, the emission radiated at some energy $ E $ (per unit energy interval) is calculated as
\begin{equation}
\label{eq:demspec}
L_E(E)=\int_{T_{min}}^{T_0}d\ln T~DEM(T)~\int dE'~K(E,E')\epsilon_{Z}(E',T),
\end{equation}
with the plasma emissivity function at specific energy $ \epsilon_{z}(E,T) $ being calculated also by means of the \texttt{APEC} model.

Convolution with the kernel $ K(E,E')$ describes spectrum smearing due to the residual velocity field present in a slice of constant temperature, which is assumed to be the same along the jet as long as deviations from ballistic flow pattern and optical depth effects are negligible. For small opening angles $ \Theta $, the redistribution kernel can be approximated as a Gaussian with $ \Delta E_{FWHM}/E=\sqrt{3}~\Theta~\gamma\beta\sin\phi $, where $ \gamma=1/\sqrt{1-\beta^2} $ and $ \phi $ is the angle between the jet axis and the line of sight \citep{Marshall2002}. The smearing thus depends on the observing conditions, and it is convenient to take it into account upon the main calculation. Hence, in what follows, we focus on $ \mathcal{L}_E(E)=\int_{T_{min}}^{T_0}d\ln T~DEM(T)\epsilon_{Z}(E,T)$, with $ L_{E}=\mathcal{S}\ast \mathcal{L}_E $, where $\mathcal{S}$ stands for the smearing operator. Similarly, spectrum transformation due to the transition from the comoving to the observer's reference frame is imposed at a later stage (see Section \ref{ss:xspec}). Contributions of different parts of the jet to the net emission are illustrated in Fig.\ref{f:speccontr}.
\begin{figure}
\centering
\includegraphics[width=1.1\columnwidth]{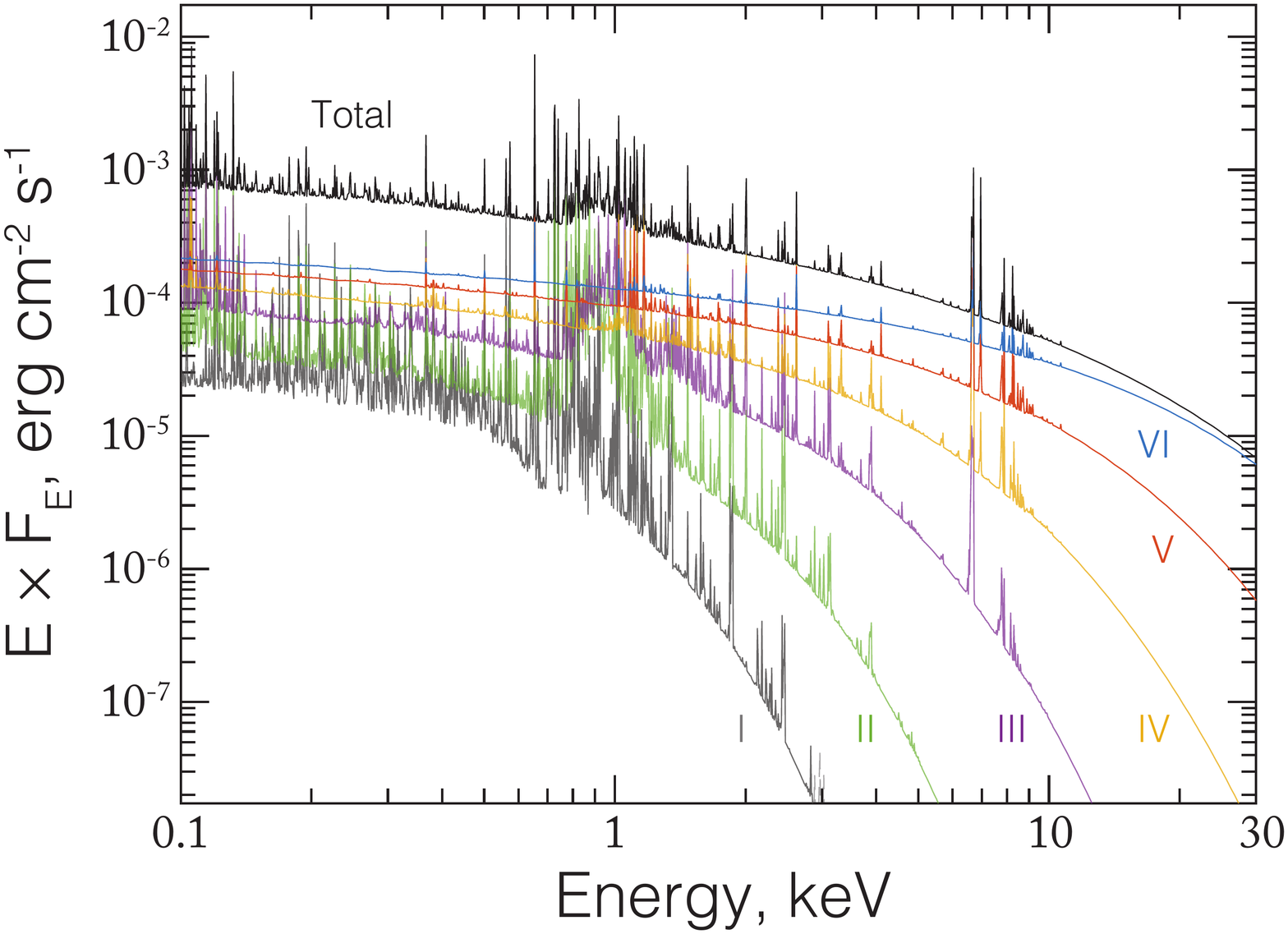}
\caption{Contribution of different portions of the jet to the net calculated emission (topmost black curve) for the model with $ T_0=20$ keV, $ \alpha=0.5 $ and solar abundances of heavy elements. The jet slices are equally-spaced on a logarithmic scale of temperature from 0.2 to 20 keV with a step factor $ 10^{1/3}\approx 2.154$, so the curve labelled \textbf{I} corresponds to $T \in [0.2,~ 0.431]$ keV, ~ \textbf{II} -- $[0.431,~ 0.928]$, ~ \textbf{III} -- $[0.928,~ 2]$, ~  \textbf{IV} -- $[2,~ 4.31]$,  ~ \textbf{V} -- $ [4.31,~ 9.28]$, ~ ~ \textbf{VI} -- $[9.28,~ 20]$ keV .
}

\label{f:speccontr}
\end{figure}

The dependence of the calculated emission on physical parameters of the jet can also be divided into a dependence of the net X-ray luminosity (or normalisation) $\mathcal{L}_{0}=\int dE \mathcal{L}_E(E)$, and a dependence of the reduced spectral shape $ \varphi(E)=\mathcal{L}_E(E)/\mathcal{L}_{0} $. Actually, it is $ \mathcal{L}_{0} $ that varies most significantly as a function of jet parameters. Indeed, it is easy to find that
\begin{equation}
\label{eq:lum}
\mathcal{L}_{0}=\pi n_{e0}^2 r_{0}^3\Theta^2\Lambda_{Z}(T_0)~\int_{\eta_{min}}^{1}d\ln\eta~\frac{\lambda_{Z,T_0}(\eta)}{\xi(\eta)}\abs{\frac{d\ln\xi}{d\ln\eta}}.
\end{equation}
 {By means of Eq.(\ref{eq:demunitless}), Eq.(\ref{eq:demad}) and  Eq.(\ref{eq:demrad})}, this is elaborated into
\begin{equation}
\label{eq:lumad}
\mathcal{L}_{0}=\frac{3\pi}{4} X n_{e0}^2 r_{0}^3\Theta^2\Lambda_{Z}(T_0)\int_{\eta_{min}}^{1}d\eta\frac{\lambda_{Z,T_0}(\eta)}{\eta^{1/4}}\propto n_{e0}^2 r_{0}^3\Theta^2\Lambda_{Z}(T_0)
\end{equation}
and 
\begin{equation}
\label{eq:lumrad}
\mathcal{L}_{0}=\pi\frac{X n_{e0}^2 r_{0}^3\Theta^2}{\alpha} \Lambda_{Z}(T_0)\int_{\eta_{min}}^{1}d\eta\frac{\lambda_{Z,T_0}(\eta)}{\eta^{1/2}}\propto n_{e0} r_{0}^2\Theta^2 \beta~T_0
\end{equation}
for an adiabatic ($ \alpha \ll 1 $) and radiatively-dominated ($ \alpha\gg 1 $) jet respectively. On the other hand, $ \varphi(E) $ is determined mainly by the \textit{shape} (and the starting point) of the DEM(T) distribution, which varies rather gently with change of the parameters' values (see Fig.\ref{f:demprof}). This is extremely useful for the technical implementation of the model, since the latter is essentially based on interpolation of spectra calculated for some (preferably not very large) grid of the parameters' values (see Section \ref{ss:xspec}). 

\subsection{Triplets of Helium-like Ions  }
\label{ss:helike}
Ratios of components in triplets of helium-like ions of heavy elements are commonly used as diagnostic tools for physical conditions in hot optically thin plasma (see \cite{Porquet2010} for a review). An attempt to take advantage of them in application to SS 433 jets was first made by \cite{Marshall2002}, and then repeated by \cite{KS2012} and \cite{Marshall2013}. Here, our aim is to allow the model to predict these ratios as a function of jet parameters in a  significantly more self-consistent (with the overall calculated emission) manner. 

While the temperature dependence of the $ G(T)=(i+f)/r $ ratio (hereafter $r,i,f $ stand for intensities of resonant, inter-combination and forbidden components, respectively) is automatically accounted for by the \texttt{APEC} model, the density dependence of the  $ R(n_e)=f/i $ ratio due to damping of the forbidden component into the intercombination one is not incorporated in it. To improve on this aspect, we extracted $ R(n_e) $ ratios\footnote{ In fact, $ R(n_e) $ is also slightly dependent on $ T $ : $ R(n_e,T)=R_0(T)/(1+n_e/n_{e,c}(T)) $, $ R_0(T)$ is the value of $ R $ in the low-density limit, and the critical density $ n_{e,c}(T) $ marks the point when the collisional de-excitation time from the upper level of the forbidden line equals the corresponding radiative transition time (e.g. \cite{Porquet2010}).} for those ions from the \texttt{CHIANTI} database (version 7.1.4, \cite{Dere1997,Landi2013}) that are most relevant for the problem in hand. We then used these ratios to correct the intensities of forbidden and intercombination lines predicted by the \texttt{APEC} model for a given portion of the jet. This procedure was performed for helium-like neon ($ E_f=0.905 
 $ keV, $ E_i\approx0.915 $ keV\footnote{An intercombination line is
  in fact a relatively close doublet (see e.g. \cite{Porquet2010}),
  therefore the cited line positions correspond to the mean of the two
  components for all intercombination lines.}, $ E_r=0.922 $ keV),
silicon ($ E_f=1.839 $ keV, $ E_i\approx1.854 $ keV, $ E_r=1.865 $
keV), sulphur ($ E_f=2.430 $ keV, $ E_i\approx2.447 $ keV, $
E_r=2.4601 $ keV) and iron ($ E_f=6.636 $ keV, $ E_i\approx6.670 $
keV, $ E_r=6.700 $ keV) (see also a more detailed list in
\cite{KS2012}). These lines are the most promising for diagnostic
purposes, since they are expected to be sufficiently bright under the
circumstances we are interested in and probe the useful range of
electron densities \citep{Porquet2010}. 

Because jet emission in a given line originates for some range of
temperatures and densities (see e.g. Fig.4 in \cite{KS2012}), drawing
any direct conclusions regarding the jet parameters from the line
ratios is fairly challenging (see Section \ref{s:ss433} and
Fig.\ref{f:triplets} there), and they actually are intended to be used
only in combination with the overall predicted emission.  {One
  should also be aware of the fact that the $ R $ ratio is sensitive
  to UV illumination of the radiating gas, owing to the possibility of
  excitement of an electron from the upper level of the forbidden
  transition to the upper level of the intercombination transition by
  a UV photon \citep{Porquet2001}. This might be of particular
  relevance to SS 433-like sources \citep{KS2012,Marshall2013},
  possessing an UV-bright supercritical accretion disc
  \citep{Gies2002}.} 
 
\subsection{Model limitations and parametrisation}
\label{ss:param}

The set of jet parameters exploited above $\mathbb{P}_0=(r_0, T_0, n_{e0}, \Theta, \beta, Z )$ not only provides a fairly natural description of the physical model, but also allows a lucid calculation of the emergent X-ray spectrum. However, it turns out that both its normalisation and shape are actually determined almost solely by some combinations of these parameters or their functions (like $ \alpha $ and $ \mathcal{L}_{0} $). Besides that, there must be some limitations on the parameter values arising from validity of the assumptions made. Finally, since technical implementation of the model is based on interpolation of spectra calculated on some sufficiently dense grid of parameter values, it is important to construct such a grid that it would cover the relevant domain of the parameter space as efficiently as possible (e.g. with minimal number of irrelevant points). Below we introduce a modification to the original set of parameters aimed to address these issues.            

A crucial assumption of the previous consideration is that every jet slice is optically thin. This makes it possible for the radiative cooling term in the thermal balance equation Eq.(\ref{eq:cooling}) to be formulated in terms of the optically thin emissivity function $ \Lambda_{Z}(T)$, so that the net spectrum can be found by integration over the corresponding DEM distribution (Eq.(\ref{eq:demspec})). The validity of this assumption is determined by the \textit{transversal} optical depth at the jet base with respect to electron scattering
\begin{equation}
\tau_{e0}=n_{e0}~\sigma_{e}~\Theta r_0, ~~ \sigma_{e}=\sigma_{Th}=6.65\times 10^{-25}~cm^2 
\label{eq:tau}
\end{equation}   
since we are safely in the Thomson limit ($ E, T_e\ll m_e c^2 $). Obviously, the transversal optical depth decreases as $ 1/r $, so the condition $\tau_{e0} < 1 $ at the jet base ensures that $\tau_{e}$ is less than unity everywhere along the jet. Moreover, even if there is a visible part of the jet with $\tau_{e0} >1 $, its emission will be suppressed by a factor of $ ~\tau_{e} $ since only a photosphere of width $ R/\tau $ will actually contribute to the observed emission from such a slice
. In this case, one can treat the point where $ \tau_e\sim 1 $ as an actual jet base. Nonetheless, the scattering cross-section for line photons is much larger due to resonant scattering even taking into account the residual velocity gradient present in the jet slices \citep{KS2012}. Resonant scattering, however, affects mainly the fine structure of the profiles of the brightest lines of highly abundant elements like Si,S and Fe, whereas a decrease in the line flux due to the combined operation of electron and resonant scattering proves to be significant only for some special configurations of the jet parameters (e.g. for a dense and compact or almost cylindrical jet, see \cite{KS2012}) \footnote{Of course, using the transversal optical depth is correct only for a jet strictly perpendicular to the line-of-sight. However, the difference becomes significant only for small viewing angles of order of $ \sin^{-1}(\tau_{e0}) $. In such a case, the spectrum can be distorted more strongly by electron and resonant scattering, due to the smaller velocity gradient along the line of sight.}. 
 {In the context of SS 433, \cite{Kotani1996} pointed out that $
  \tau_{e0}$ should be less than unity since no broad Compton wings of
  the lines were observed in the spectra measured by \textit{ASCA},
  which was later confirmed by \cite{Marshall2002} with the aid of
  \textit{Chandra}/HETGS data.}  
We neglect the opacity effects in this work, but impose a somewhat conservative limit $ \tau_{e0}\leq 0.5 $, which is also dictated by the significantly degrading accuracy of model interpolation between points with large $ \alpha $ (see Appendix), while
\begin{equation}
\alpha =  \frac{2}{3} \frac{\tau_{e0}}{\Theta\beta}~\frac{\Lambda_{Z}\left(T_{0}\right)}{\sigma_e c~T_{0}} \frac{X}{1+X}
\label{eq:alphatau}
\end{equation}
 thus $ \alpha \propto \tau_{e0} $ for fixed $ \Theta, \beta $ and $ T $ . The latter relation also demonstrates that the dependence of $ \alpha $ on $ n_{e0}$ and $r_0 $ actually reduces to a dependence on $ \tau_{e0} $. Moreover, one has
\begin{equation}
\alpha \approx  23.0~\tau_{e0} \times \left(\frac{0.01}{\Theta}\right)~\left(\frac{0.1}{\beta}\right)~\left(\frac{10~keV}{T_{0}}\right)^{1/2}
\label{eq:alphafid}
\end{equation}
 {for radiative cooling determined by hydrogen and helium bremsstrahlung only.}
Therefore, small $ \tau_{e0}\lesssim 0.01 $ also implies small $ \alpha$, and hence there is no need to calculate spectra for very low  $ \tau_{e0} $ because the spectral shape is invariant in the low-$ \alpha $ regime.

As regards the model normalisation, it is natural to expect that $ \mathcal{L}_{0} $ is somehow related to the available energy budget of the jet. Indeed, the equation for the model normalisation $ \mathcal{L}_{0} $  in the high-$ \alpha $ regime (Eq.(\ref{eq:lumrad})) suggests that it is simply the flux of thermal energy through the jet 
\begin{equation}
L_{th}=\pi ~ r_0^2 \Theta^2~(n_{e0}+n_{i0})~\beta~c~\frac{3}{2} T ,
\label{eq:lth}
\end{equation}
while in the low-$\alpha $ regime one has $\mathcal{L}_{0}\sim \alpha
L_{th} $ (see Eq.(\ref{eq:lumad})). Thus, $ \mathcal{L}_{0} $ scales
linearly with $ L_{th} $ for a fixed $ \alpha $, and $
\mathcal{L}_{0}\leq L_{th}$ in all cases. The thermal energy flux,
however, constitutes only a minor fraction of  {a mildly
  relativistic} jet's energy budget, which is dominated by the kinetic
power of the bulk motion, 
\begin{equation}
L_k=(\gamma-1)\dot{M_{j}}c^2\simeq\frac{\beta^2}{2}\dot{M_{j}} c^2=\pi ~ r_0^2 \Theta^2~(n_{e0}+n_{i0})~\beta~c~\frac{~\beta^2}{2} \mu m_{p}c^2,
\label{eq:lkin}
\end{equation}    
 {if the jet's plasma is not very hot:}
\begin{equation}
L_{th}/L_{k} = \frac {3T}{(\gamma-1)\mu m_pc^2}\simeq \frac{10T_0}{\beta^2 m_pc^2} \sim 10^{-2}~\times~\left(\frac{T}{10 ~ keV}\right)\left(\frac{\beta}{0.1}\right)^{-2}
\label{eq:lthlk}
\end{equation} 
for $ \gamma=1/\sqrt{1-\beta^2}\approx 1-\beta^2/2 $ and $ \mu \approx 0.6 $.

 Being large enough both in absolute and relative senses, $L_k$ has two major advantages. First, since the kinetic power of the jet dissipates very slowly, staying almost constant up to the termination point of the jet, it can be measured at various scales independently of X-ray observations of the central object (e.g. from the energy deposited in the circumjacent interstellar medium). Second, if the kinetic luminosity of the jet indeed dominates the energy output of the accreting compact object, one may expect it to be comparable but not much higher than the Eddington luminosity, given that accretion proceeds in the super-critical regime (see results for Model A in \cite{Ohsuga2011}). Hence, there is a natural upper limit for this quantity of order of $ 10^{40} $ erg/s as far as we are dealing with jets in XRBs. On the other hand, taking into account Eq.(\ref{eq:lthlk}) and recalling that $ \mathcal{L}_{0}\leq L_{th}$ in all cases, we can easily find a lower limit for $ L_k $ provided that $ \mathcal{L}_{0}$ and $ T_0 $ are known:
\begin{equation}
L_{k}\geq \mathcal{L}_{0}~\frac{\beta^2~m_pc^2}{10~T_0}\sim 100~\left(\frac{10~keV}{T_0}\right)\left(\frac{\beta}{0.1}\right)^{2}\times \mathcal{L}_{0}. 
\end{equation}
This immediately puts a constrain on the minimum $ L_{k} $ we may be interested in, given typical X-ray flux limits for spectroscopy of sufficient quality. Say, for $ f_x=10^{-14} $ erg/s/cm$^{2}$, $ \mathcal{L}_{X0,min}\sim 10^{30}~d_{kpc}^2 $ erg/s, i.e. $ {L}_{k}\gtrsim 10^{32}~d_{kpc}^2 $ erg/s. Additionally, as noticed before, $\alpha \sim \tau_{e0}$ for given $ \beta $, $ \Theta $ and $ T_0 $, whereas $\mathcal{L}_{0}\propto \alpha L_{k} $ and the spectral shape is invariant in the low-$ \alpha $ regime. Therefore, there is degeneracy between $ \tau_{e0} $ and $ L_k $ for small $ \tau_{e0} $, since the model prediction is sensitive to their product only. In principle, this degeneracy may be broken by means of density-sensitive line ratios in He-like triplets (see Section \ref{ss:lineratios}).  

As a result, being naturally constrained and having understandable influence on characteristics of the predicted emission, both $ \tau_{e0} $ and $ L_{k} $ are well suited for more efficient model parametrisation. Therefore, $ r_0 $ and $ n_{e0} $ can be replaced by the pair of new parameters, $ L_{k} $ and $ \tau_{e0} $. In order to perform the corresponding transformation, it is useful to rewrite $ L_k $ using more appropriate units.

 As mentioned above, a natural scale for $L_{k}  $ is set by the Eddington luminosity, which for a compact object with of mass $ M $ is 
\begin{equation}
L_{E}=\frac{4\pi G c (1+X)\mu m_p}{\sigma_{e}} M =L_{E,\odot}\frac{M}{M_{\odot}},
\label{eq:led}
\end{equation}
$L_{E,\odot}\approx 1.489\times 10^{38}$ erg/s. As follows from Eq.(\ref{eq:lkin}),
\begin{equation}
\frac{L_{k}}{L_{E}}=~\frac{\sigma_{e}~c^2}{8GM}~~n_{e0}~ r_0^2~\Theta^2~\beta^3 =\frac{\sigma_{e}}{4~r_{g}} ~~n_{e0}~ r_0^2~\Theta^2~\beta^3
\label{eq:lkled}
\end{equation}
where $ r_g=2GM/c^2=r_{g,\odot}~M/M_{\odot}\approx 3\times 10^5 ~M/M_{\odot}$ cm.
Dividing this by $ \tau_{e0} $, results in 
\begin{equation}
{r_0}=4r_{g}\frac{{L_{k}}/{L_{E}}}{\tau_{e0}\beta^3\Theta} ~\approx 1.2\times 10^{6}~cm ~~ \frac{{L_{k}}/{L_{E,\odot}}}{\tau_{e0}\beta^3\Theta}
\label{eq:r0lkled}
\end{equation}
and 
\begin{equation}
n_{e0}=\frac{1}{4r_g\sigma_e}~~\frac{\tau_{e0}^2\beta^3}{{L_{k}}/{L_{E}}}\approx 1.25\times 10^{18}~cm^{-3}~~\frac{\tau_{e0}^2\beta^3}{{L_{k}}/{L_{E,\odot}}}.
\label{eq:n0lkled}
\end{equation}
These relations describe the transformation from $\mathbb{P}_0=(r_0, n_{e0}, T_0, \Theta, \beta, Z $) to $\mathbb{P}=( L_{k}, \tau_{e0}, T_0, \Theta, \beta, Z) $, and it is the latter set of parameters that is actually used in our implementation of the model. 

With the new set of variables, Eq.(\ref{eq:lumad}) implies that
\begin{equation}
\mathcal{L}_{0}(\mathbb{P})=C(\mathbb{P}) ~\frac{\tau_{e0}L_{k}}{\theta\beta^3} 
\label{eq:lx0ad}
\end{equation}
where $C(\mathbb{P})$ depends on $ \mathbb{P} $ only weakly for the low-$ \alpha $ regime. 

For the high-$ \alpha $ regime (cf. Eq.(\ref{eq:lumrad})), one has
\begin{equation}
\mathcal{L}_{0}(\mathbb{P})=C(\mathbb{P}) ~\frac{L_{k}~T_0}{\beta^2}. 
\label{eq:lx0rad}
\end{equation}

 {Note that there might exist a physical relationship between
  the parameter values, such as one suggested by \cite{Marshall2002}
  for $ \Theta $, $ \beta $ and $ T_0 $: $ \Theta~\beta c\sim
  c_{s}=\sqrt{\frac{5~T_0}{3(1+X)\mu m_p}} $, i.e. $ \beta\Theta \sim
  \sqrt{T_0/m_{p}c^2} $, meaning that the jet expands sideways at the
  sound speed of plasma at the jet's base. Such constraints may be
  easily taken into account by setting corresponding links between the
  parameters when applying the model (e.g. in \texttt{XSPEC}).}  

 {We finally note that the X-ray emitting jet is assumed to be
  a continuous flow in our model, while one might also be interested in
  considering a jet consisting of $ N $ distinct clumps or streamlets
  with a mean opening angle $ \theta$ (see e.g. \cite{Koval1989} and
  \cite{Atapin2015}). In this case, for the same $ \beta $ and
  \textit{net} $ L_{k} $, each clump will be characterized by an
  initial transversal optical depth
  $\tilde{\tau}_{e0}=\tau_{e0}\frac{\Theta}{N\theta}$. Therefore, if  
  $\tilde{\tau}_{e0}$ is sufficiently small and $ \theta $ is within
  the allowed range, the cumulative emission of the jet can be
  estimated by applying our model to a single streamlet (i.e. for $
  \mathbb{P}=(L_{k}/N,\tilde{\tau}_{e0}, T_0, \theta, \beta, Z)$) and
  artificially increasing the normalisation and spectral smoothing
  width (see Section \ref{ss:xspec} and Appendix) by a factor of $N$.}    

\section{Basic predictions}
\label{s:observ}
%
In this section, we analyse the basic predictions of the model and probe their sensitivity to the model parameters across the potential range of the parameter values. Some boundaries of this range have been already outlined above (namely, for $\tau_{e0} $ and $ L_k $), while some others are discussed in the Appendix. Here we just mention that $ L_{k} $ is allowed to vary from $ 10^{35} $ to $ 10^{41}$ erg/s, $\tau_{e0} $ from $ 5\times10^{-5} $ to 0.5, $ T_0 $ from 7 to 40 keV, $~ \Theta $ from 0.003 to 0.03 rad, $~ \beta $ from 0.03 to 0.3 and $ Z $ from 0 to 9.

\subsection{Continuum and line luminosities}
\label{ss:norm}

First of all, we are interested in the range of jet X-ray luminosities available to the model. 
As shown in the top panel of Fig.\ref{f:lum}, the model $ \mathcal{L}_{0} $ covers a range from $\sim 10^{28} $ to $\sim 10^{40} $ ergs/s for each value of $ T_0 $. This range is mainly provided by the $ \frac{\tau_{e0}L_{k}}{\theta\beta^3} $ factor in Eq.(\ref{eq:lx0ad}), so dividing $ \mathcal{L}_{0} $ by this factor reduces the range to a relatively narrow stripe (see Fig.\ref{f:lum}). This makes it possible to increase the accuracy of table model interpolation (see Appendix).

Next, we consider the contributions of different energy bands to the total luminosity. We divide the spectrum in three broad bands: soft (E$ < $1.5 keV), medium (1.5 keV$ < $E$ < $9 keV ) and hard (E$ > $9 keV). The relative contributions of these bands are shown in the second top panel of Fig.\ref{f:lum} for models with $ Z=1$. Noteworthy, the contribution of the medium band varies only slightly across the parameter space, and it equals $ \approx 0.4 $ everywhere. As expected, the contribution of the soft band decreases with increasing $ T_0$, and the opposite is true for the hard band. Since the soft band is heavily affected by interstellar absorption while the hard band is beyond the effective sensitivity range of the \textit{XMM-Newton} and \textit{Chandra} observatories, we focus more on the medium band.

Namely, we introduce three sub-bands: 1.5-3, 3-6 and 6-9 keV. This is motivated by the fact that there are no strong lines in the 3-6 keV energy sub-band and the emitted spectrum can be described in terms of a simple power-law model there. On the other hand, emission lines contribute singificantly to the flux from 1.5 to 3 keV (mainly lines of H- and He-like silicon and sulphur) and above 6 keV (due to H- and He-like iron). Again, fluxes in the softer (1.5-3 keV) and harder (6-9 keV) sub-bands are fairly sensitive to the DEM of relatively cold ($ T\sim 1 $ keV) and hot ($ T\gtrsim 10 $ keV) gas, while the contribution of the 3-6 keV sub-band to the total luminosity remains almost constant at a level of $ \approx 0.14 $ for $ Z=1 $ (see the second top panel of Fig.\ref{f:lum}). 

The dependence of luminosities in the strongest lines of the softer and harder sub-bands on $ T_0 $ is demonstrated in the two bottom panels of Fig.\ref{f:lum}. The major contribution comes from the Fe XXV K$\alpha $ triplet line ($ \sim 1\%$), while the other lines provide typically $ \sim few\times 0.1\% $ of the total luminosity for $ Z=1 $ (and hence 2.5 times more to the 1.5-9 keV luminosity). Of course, these numbers scale almost linearly with $ Z $, and below we consider how this (along with variation in other parameters) affects the spectral energy distribution of the model in the medium band.

\begin{figure}
\centering
\includegraphics[bb=40 170 560 690,width=1.\columnwidth]{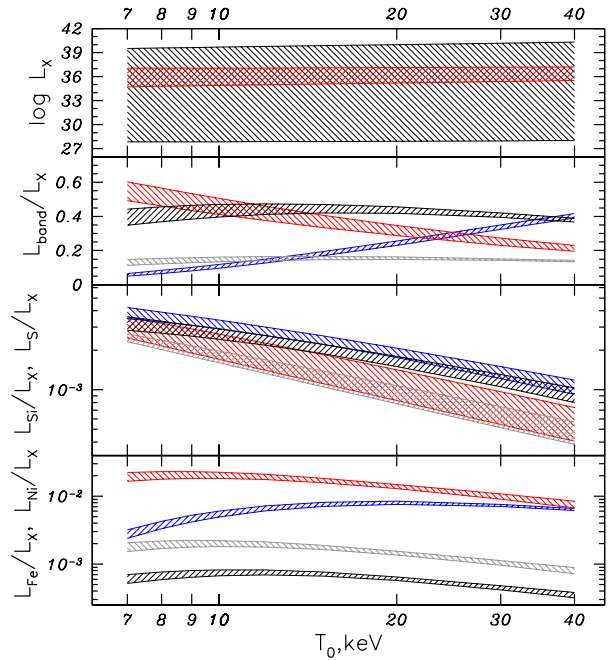}
\caption{\textit{Top panel.} The range of total X-ray luminosities (0.1-100 keV) covered by the model with $ Z=1 $ is dashed with black lines. The region dashed with red lines demonstrates the effect of model re-normalisation by dividing by $N = \tau_{e0} L_{38}/\Theta_{0.01}\beta_{0.1}^{-3} $, where $L_{38}=L_{k}/10^{38}$ erg/s, $\Theta_{0.01}=\Theta/0.01$ and $ \beta_{0.1}=\beta/0.1$ (see Section~\ref{ss:norm}). \textit{Second top panel.} Dependence of the contribution from the soft ($ < $1.5 keV, red), medium (1.5-9 keV, black) and hard ($ > $9 keV, blue) energy bands to the total X-ray luminosity on $ T_0 $ (for $ Z=1 $). The contribution of the 3-6 keV energy band is shown in gray. \textit{Third top panel.} Dependence of the contribution from the strongest lines in the 1.5-3 keV energy band on $ T_0 $ (for $ Z=1 $): Si XIII K$ \alpha$ triplet -- red, Si XIV Ly$ \alpha$ -- blue, S XV K$ \alpha$ triplet -- gray and S XVI Ly$ \alpha$ -- black. \textit{Bottom panel.} Dependence of the contribution from the strongest lines in the 6-9 keV energy band on $ T_0 $ (for $ Z=1 $): Fe XXV K$ \alpha$ triplet -- red, Fe XXVI Ly$ \alpha$ -- blue, Fe XXV K$ \beta$ -- gray, Ni XXVII K$ \alpha$ triplet -- black .  }
\label{f:lum}
\end{figure}
\subsection{3-6 keV slope, $ F_{1.5-3}/F_{3-6}, ~F_{6-9}/F_{3-6} $ and Ly$\alpha$/ K$ \alpha $ flux ratios}
\label{ss:slopenratios}

The power-law spectrum in the 3-6 keV energy sub-band arises from a combination of bremsstrahlung emission from various parts of the jet, and its slope should be determined mainly by the maximum visible temperature, since the $ DEM$ decreases with decreasing $ T $ (see Section \ref{ss:speccalc}). Indeed, the dependence of the power-law slope $ \Gamma $ on $ T_0 $ is clearly seen in the top panel of Fig.\ref{f:observ}. Increasing metallicity makes the spectrum in this region somewhat harder, as a result of suppression of the $ DEM $ of low-temperature parts of the jet (see Fig. \ref{f:demprof}), and  increases the scatter in the $ \Gamma-T_0 $ relation caused by other parameters.   

As the spectrum gets harder, the ratio of fluxes in the 6-9 keV to 3-6
keV sub-bands must grow correspondingly, and this effect is
demonstrated in the second top panel of
Fig.\ref{f:observ}.  {However, as noticed above, there is a
  significant contribution of line emission (primarily due to H- and
  He-like iron) to the spectrum in the 6-9 keV sub-band, so the
  $F_{6-9}/F_{3-6}  $ ratio depends more on the equivalent width of
  the lines, which is sensitive to the gas metallicity $ Z $ since
  the continuum emission is dominated by bremsstrahlung from hydrogen
  and helium (see Fig.\ref{f:observ}).} This, in combination with a
measurement of the slope in the 3-6 keV sub-band, allows one to
estimate the required metallicity just from the flux ratios with no
need for resolving the lines. The scatter caused by other parameters
is vanishingly small, and it becomes noticeable only for extremely
high values of $ Z $.  

Although there is also a significant contribution from lines to the flux in the 1.5-3 keV energy sub-band, the dependence of the $F_{1.5-3}/F_{3-6}  $  ratio on metallicity is not so clear in this case (see the third top panel of Fig.\ref{f:observ}), and the main effect of increasing metallicity is actually an increased sensitivity to other model parameters, as reflected by a large scatter in the $F_{1.5-3}/F_{3-6}(T_0, Z) $ relation. This also implies that fitting the spectrum in this energy sub-band may provide the most stringent constrains on these additional parameters, given that $ T_0 $ and $ Z $ can be estimated from data in the harder sub-bands only.    

Additionally, there is a well-known temperature diagnostic tool based on intensity ratios of the K$\alpha$ triplet line of He-like ions to the Ly$\alpha$ line of H-like ions. Only modest spectral resolution is required to resolve these lines. In the context of SS 433 jets, this technique was first exploited by \cite{Kotani1996} for iron lines in spectra provided by the $ ASCA $ observatory and was then extensively used for \textit{XMM-Newton} (e.g. \cite{Medvedev2010}) and \textit{Chandra} (e.g. \cite{Marshall2002}) data. In the bottom panel of Fig. \ref{f:observ}, we show the dependence of the Ly$\alpha$/K$\alpha$ flux ratio on $ T_0 $ for iron, sulphur and silicon. Since we consider only $ T_0>7 $ keV, this ratio is always higher than unity for sulphur and silicon and its dependence on $ T_0$ is modest and comparable to the dependence on other parameters, changing the DEM shape at $ T\sim 2 $ keV. Contrary to this, the Ly$\alpha$/K$\alpha$ ratio for iron is indeed an excellent diagnostic of $ T_0 $ with very small scatter caused by other parameters. There is also a finer diagnostic based on the Fe XXV K$ \beta $/Fe XXV K$ \alpha$ flux ratio, but the former line is typically ten times fainter than the latter, and the relative variations in this ratio are only of order 10\%, so taking advantage of it is hardly feasible.  

\begin{figure}
{\centering
\includegraphics[bb=40 200 560 690,width=1.\columnwidth]{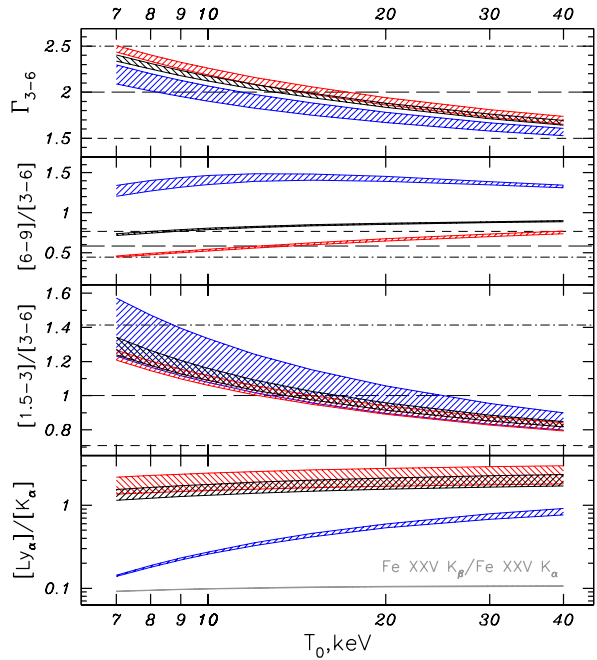}
}
\caption{Predictions for basic features of the calculated spectra as a function of the jet base temperature $T_0 $ and metallicity with the scatter coming from dependence on the other parameters. \textit{Top panel.} The slope of a power-law fit to the spectrum in the 3-6 keV energy band. The black, red and blue regions correspond to $ Z=1 $, $ Z=Z_{min}=0 $ and $ Z=Z_{max}=9 $, respectively. The horizontal lines mark $ \Gamma=1.5 $ (short-dashed), $ \Gamma= 2$ (long-dashed) and $ \Gamma=2.5 $ (dot-dashed).  \textit{Second top panel.} The same as above, but for the [6-9 keV]-to-[3-6 keV] flux ratio. The horizontal lines demonstrate the corresponding values expected for power-law spectra with $ \Gamma=1.5 $ (short-dashed), $ \Gamma= 2$ (long-dashed) and $ \Gamma=2.5 $ (dot-dashed). \textit{Third top panel.} The same as above, but for the [1.5-3 keV]-to-[3-6 keV] flux ratio. \textit{Bottom panel.} Dependence of the flux ratio of the Ly$ \alpha$ line of the H-like ion to the K$ \alpha$ line of the He-like ion on $ T_{0}$ ($Z=1$) for Si (red), S (black) and Fe (blue). Also,  the FeXXV K$\beta $/FeXXV K$\alpha $ flux ratio is shown in gray. }
\label{f:observ}
\end{figure}
\subsection{Line ratios in He-like triplets}
\label{ss:lineratios}

Although theoretically the intensity ratio of the forbidden line to the inter-combination one (the $R$-ratio) provides a rather direct way to measure the number density of the emitting gas, in reality one has to deal with a number of complications. First of all, plenty of fainter neighbouring lines blend with the actual triplet components due to the finite energy resolution of any spectrometer. Moreover, there is broadening of the lines intrinsic to the model itself. In our case, this comes from the conical flow pattern and is determined by the jet opening angle $ \Theta $.  

Secondly, since we consider the multi-temperature model, the actual $R$-ratio is a weighted mean over the jet (with the weighting proportional to the DEM) and is thus somewhat biased to higher densities and also corresponds to some higher temperature $ T_{eff} $ than the temperature of peak emissivity. Fortunately, the dependence of the $R$-ratio on effective temperature is relatively weak, but the situation may be different for some satellite lines. One of the major contributors to satellite lines is dielectronic recombination, which results in lower peak temperatures for the satellites than for the triplet components. Again, since the DEM is decreasing with decreasing temperature, the influence of these satellites is expected to be relatively suppressed. However, there are also satellite lines having higher peak temperatures. This is particularly the case for the intercombination line of neon and the forbidden line of silicon, for which the contribution of the Fe XIX and Mg XII Ly$ \gamma $ lines, respectively, is important.          

Finally, as the model parameters define the number density $ n_{e0} $ at the jet base, the actual density at the position where the triplet emission mostly comes from (roughly, the point with temperature equal to the effective one) differs from $ n_{e0} $, and this difference is larger for a larger difference between $ T_0 $ and $ T_{eff} $. This makes the sensitivity range exceptionally broad for low-$ T_{eff} $ elements, first of all neon and silicon, but much narrower for sulphur and iron. The situation becomes simpler in the high-$ \alpha $ regime, since the jet is 'short' here and $ n_{e} $ does not vary significantly along it, and equals $ n_{e0} $ everywhere. This almost completely eliminates the dependence on $ T_0 $ in this case.

Of course, all these effects are automatically accounted for when the model is fitted locally in the triplet region. Instead, here we try to provide a  qualitative picture of the sensitivity of the $R$-ratio to the model parameters. For simplicity, we fixed the spectral smearing width at the resolution of \textit{ Chandra} HEG, i.e. $ \delta\lambda=0.012\AA$ (cf. also \citep{Porquet2001}). Namely, all satellite lines in the range $ [E_0-\delta\lambda/2,E_0+\delta\lambda/2 ]$ contribute to the flux of a triplet component with centroid energy $ E_0$. The resulting picture is shown in Fig.\ref{f:triplets} for three values of $ T_0$ (10, 20 and 40 keV) and $ \alpha$ (0.01, 0.5 and 5), where all the basic features described above are clearly seen. The fact that $R$ does not go to zero in the high-density limit for silicon and iron reflects the contribution of satellites to the forbidden component. As regards the low-density regime, one may notice a difference in the limiting $R$ value for neon and silicon in the high-$ \alpha $ regime (see the top panels of Fig.\ref{f:triplets}). This is caused by the contribution of satellites with higher peak temperatures to the intercombination line of neon and the forbidden line of silicon. For the high-$ \alpha $ regime, the $DEM(T)$ profile becomes steeper for $ T<2 $ keV (see Fig.\ref{f:demprof} increasing the importance of high peak temperature satellites, correspondingly decreasing the $R$-ratio for neon and increasing it for silicon. Thus, He-like triplets indeed provide additional sensitivity of the model predictions to $ \alpha $, which is equivalent to sensitivity on $ \tau_{e0} $ for fixed $ \beta $, $ \Theta $ and $ T_0 $ (see Eq.(\ref{eq:alphatau})). In the low-$\alpha$ regime, the dependence on $ n_{e0}\propto \tau_{e0}\beta^3/L_{k} $ makes it possible to break the degeneracy of model predictions with respect to the $ \tau_{e0}L_{k}$ combination (see Section \ref{ss:param}). Below we present an example of actual use of the model in application to a high-resolution spectrum of SS 433 provided by \textit{Chandra} HETGS.

\begin{figure*}
\centering
\includegraphics[width=0.85\textwidth]{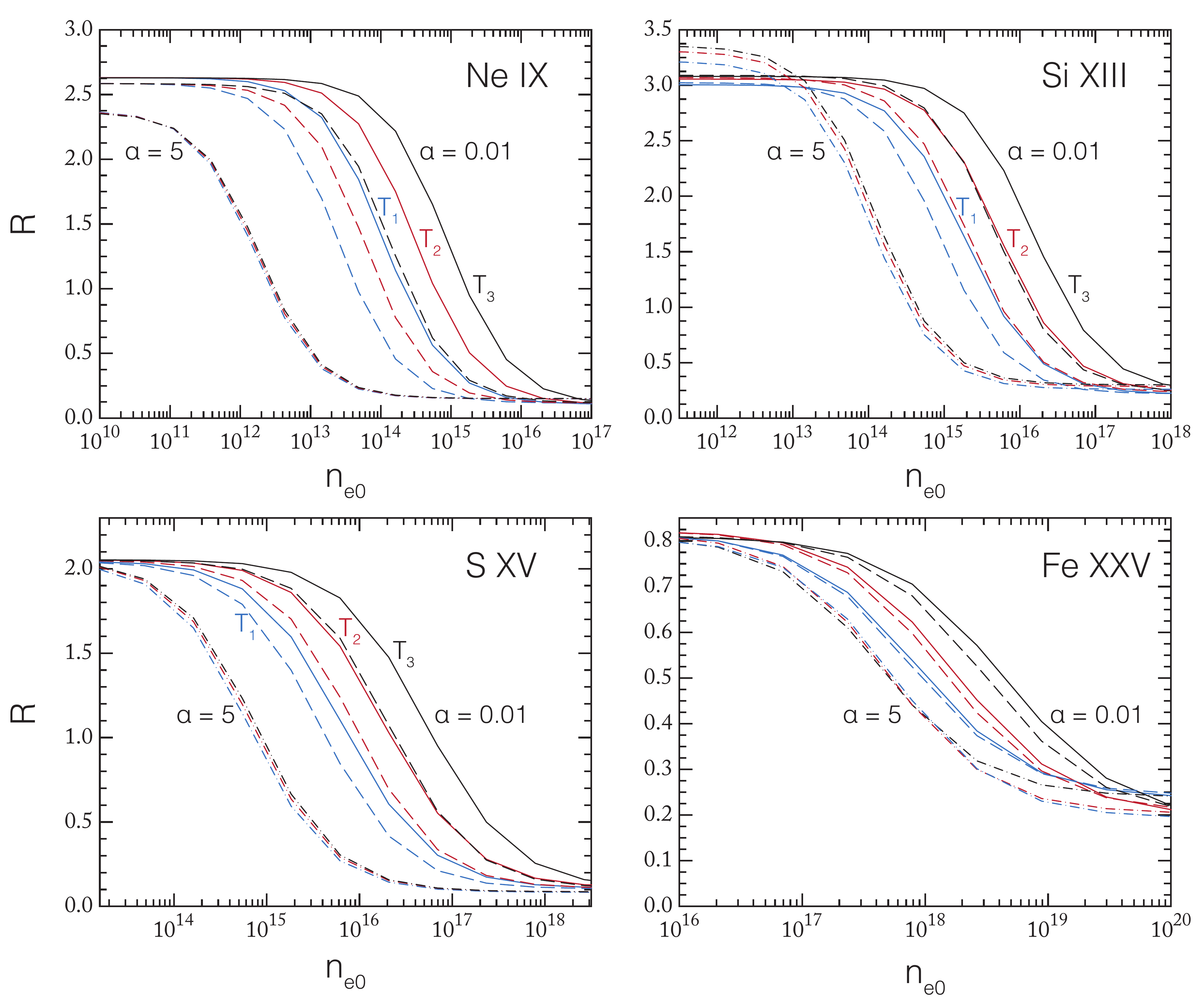}
\caption{The $R = f/i$ ratio as a function of the electron density at the jet base. The ratio is obtained from the continuum-removed spectra of the model with energy resolution corresponding to the Chandra HEG: $\delta \lambda = 0.012$ \AA. The result is given for different base temperatures, $T_1 = 10$ keV (blue lines), $T_2 = 20$ keV (red), $T_3 = 40$ keV (black), and $\alpha = 0.01$ (solid lines), $\alpha = 0.5$ (dashed), $\alpha = 5$ (dash-dotted).
}
\label{f:triplets}
\end{figure*}
\section{Application to SS 433}
\label{s:ss433}

In this section we check whether our model is indeed adequate for description of the X-ray emission of SS 433. However, since detailed analysis of SS 433 spectra (to get stringent constraints on the parameters of the system) is not the primary aim of the present paper, we focus mainly on the ability of the model to reproduce the basic features of the observed emission of SS 433. Furthermore, as has been demonstrated in previous studies, some other components may be required in addition to the thermal emission of the jets to describe X-ray spectra of this source \citep{Brinkmann2005, Medvedev2010}. In the reflection model, the observed additional component arises from emission reflected from the inner regions of the super-critical accretion disc \citep{Medvedev2010}. Indeed, the fluorescent K$ \alpha $ line of neutral iron Fe I, usually detected at 6.4 keV, strongly supports the scenario with reflection from cold matter of walls of the accretion disc funnel \citep{Medvedev2010}. However, the contribution of this component is expected to be low at the precessional phase when the accretion disc is viewed nearly 'edge-on'. We can thus expect the jet model alone to fit the data at this phase relatively well. 
 
\subsection{SS433-specific model}
\label{ss:sssmodel}

Estimates for some parameters of the model can be obtained not only by fitting X-ray spectra but also by other observational techniques. This is particularly the case for $ \beta $ and $ \Theta $, which can be determined from Doppler shifts and broadening of emission lines. Indeed, if $ z_b $ and $ z_r $ are the Doppler shifts for the approaching ('blue') and receding ('red') jets, respectively, then
\begin{equation}
\beta=\sqrt{1-\frac{1}{(1+z_0)^2}}=\sqrt{2z_0}\left(1-\frac{3}{4}z_0+\mathcal{O}(z_0^2)\right), z_0=\frac{z_r+z_b}{2}.
\label{eq:zbeta}
\end{equation}
The latter estimate supposes that both jets are perfectly aligned and have the same velocity. Some apparent deviations from the symmetrical picture were reported by \cite{Marshall2013}, which however have only a minor effect on the total predicted spectrum.

The angle $\phi $ between the axis of the jets and the line of sight may then be found either for $ z_b $ or  $ z_r $:
\begin{equation}
z_b=\gamma\left(1-\beta\cos\phi\right)-1,~~ z_r=\gamma\left(1+\beta\cos\phi\right)-1, 
\label{eq:zbzr}
\end{equation}
i.e.
\begin{equation}
\cos\phi=\frac{1}{\beta}\left(1-\frac{1+z_b}{\gamma}\right)=\frac{1}{\beta}\left(\frac{1+z_r}{\gamma}-1\right)=\frac{z_r-z_b}{2\gamma\beta}.
\label{eq:phizbzr}
\end{equation}

Further, if the profile of a line at energy $ E_0 $ is approximated by a gaussian with dispersion $ \Sigma(E_0)$, one may estimate the jet half-opening angle $ \Theta $ as \citep{Marshall2002}
\begin{equation}
\Theta=\sqrt{\frac{2\ln 2}{3}}~~\frac{2}{\beta\gamma\sin\phi}~~\frac{\Sigma(E_0)}{E_0}.
\label{eq:wtheta}
\end{equation}

Additionally, these parameters may be estimated from optical data in similar manner, assuming of course that these parameters remain constant at least up to the distances where the optical emission comes from ($ r\sim 10^{14}-10^{15} $ cm, e.g. \cite{Fabrika2004}). For SS 433, such measurements result in $ \beta\approx 0.26 $ and $ \Theta\approx0.013$ rad from X-ray data (\cite{Marshall2002}, for the correct value of $ \Theta $ see also a remark in \cite{KS2012} and  \cite{Marshall2013}). Values inferred from optical data are also broadly consistent with these estimates \citep{Borisov1987} (possibly suggesting a slight decrease in $ \beta$), although the optical lines typically exhibit relatively complex structure and possible contamination by emission from the supercritical accretion disc wind (e.g. \cite{Medvedev2013}). Therefore, by fixing $ \beta$ and $ \Theta $ at these values we can reduce the number of model parameter, while the corresponding shifting and broadening of the lines may be superimposed upon the model by means of standard spectral analysis tools (see Section \ref{ss:xspec}). 

Owing to this possibility, we may take into account another effect which is of particular relevance to SS 433. Namely, significant over-abundance of nickel was invoked to describe an observed excess of intensity at $ E\sim 7.8 $ keV (rest frame) \citep{Medvedev2010,Kubota2010,Marshall2013}, the region where the $ K\alpha $ emission line of He-like nickel is expected to be present (along with the $ K\beta $ line of He-like iron, see e.g. \cite{KS2012}). We allow nickel abundance to vary separately from other heavy elements, introducing a new model parameter $ Z_{Ni} $. Its effect on the predicted spectrum is modest and limited to the intensities of the nickel lines (see e.g. Table 4 in \cite{Marshall2013}). 
Thus, the set of parameters for SS 433-specific version of the model reads as ($L_{k}, \tau_{e0}, T_0, Z $, $ Z_{Ni} $). 

 \subsection{Data}
\label{ss:data}

The spectrum of SS 433 in the standard X-ray energy range (1-10 keV) has been extensively studied with \textit{Chandra} High Energy Transmission Grating Spectrometer (HETGS) \citep{Weisskopf2002,Canizares2005}, with the total accumulated exposure being more than 300 ks \citep{Marshall2013}. Here, we focus on an observation performed on 2001 March 16 (ObsID 1019) with a net exposure time of 24 ks \citep{Lopez2006}. The jets' inclination to the line of sight is $ \approx 80^\circ $ for this observation \citep{Lopez2006}, i.e. the accretion disk is viewed nearly edge-on. Correspondingly, \cite{Lopez2006} found $ z_b= 0.0111$ and $ z_r=0.0610 $ (with the cited uncertainty of 0.0001 in both numbers), hence $ \phi=84.72\deg $, while $\beta=0.2615  $ and $ \Theta=0.024 $ rad. This is broadly consistent with the $\beta $ and $\Theta$ values adopted for our SS 433-specific model (see above), since the value for $\Theta$ inferred from line broadening is more likely an upper limit due to possible contribution from other broadening mechanisms (like jet nutation and resonant scattering, \cite{KS2012} for a thorough discussion). These values are used as a starting point for our fitting process in the next section.    
   
The data was downloaded, prepared and reduced by means of the standard \texttt{TGCat} pipeline \citep{Huenemoerder2011} and analysis threads from the \texttt{CIAO 4.7} package. Only $ \pm 1 $ orders were combined both for the High Energy Grating (HEG) and Medium Energy Grating (MEG), but the HEG and MEG were not further combined due to significant difference in their response functions. Since a significant fraction of the spectral channels have very low signal-to-noise ratio (SNR), we used the weighting function by \cite{Churazov1996}. Besides that, we took advantage of the standard tool \texttt{grppha} from the \texttt{ftools} package \footnote{http://heasarc.gsfc.nasa.gov/docs/journal/grppha4.html} in order to produce re-binned versions of the spectra with at least 25 \textit{raw} counts per bin, for which $ \chi^2 $-statistics can be used. This allows one to check whether the derived values of the model parameters are sensitive to a particular fitting technique.  

\subsection{Model implementation and fitting in \texttt{XSPEC} }
\label{ss:xspec}

Both the approaching ('blue') and the receding ('red') jets contribute to the observed emission from the unresolved X-ray core of SS 433. We thus need a two-component model with the components appropriately Doppler shifted and boosted. In \texttt{XSPEC}, Doppler shifting may be performed using the convolution model \textit{zashift}, which also accounts for the corresponding Doppler boosting. Next, gaussian broadening of the lines is imposed by the \textit{gsmooth} convolution model, with $ \Sigma(E) $ given by Eq.(\ref{eq:wtheta}) with the values for $ \beta$, $ \gamma $ and $ \phi $ mentioned above, so $ \Sigma(E)=23.8~eV \times (E/6~keV) \times (\Theta/0.02~rad)$. Finally, interstellar absorption is accounted for in the usual fashion by the \textit{wabs} model. Thus, the full model reads as
\begin{equation}
wabs*\left(gsmooth*\left(zashift_{b}*jet_{b}+constant*zashift_{r}*jet_r\right)\right)
\label{eq:model}
\end{equation}
where the additional component \textit{constant} allows one to attenuate the red jet by some constant factor across the whole energy range (as would be in the case for obscuration by some material with gray opacities). The \textit{jet} model stands for our table model loaded with the \texttt{atable} command and with the normalisation properly determined (see Appendix). There is a dichotomy in estimates of the distance to SS 433: while \cite{Blundell2004} found $ d_{SS433}=5.5\pm 0.2 $ kpc by fitting a kinematic model to arcsec-scale radio images of the jets, estimates based on the proper motion of discrete ejecta in the inner radio jets yield $ d_{SS433}=4.6\pm 0.35 $ kpc \cite{Stirling2002}, and even lower values are derived by other less direct methods (see  discussions in \cite{Blundell2004} and more recently in \cite{Panferov2014} and references therein). In the analysis below, we consider both estimates and check the sensitivity of our results to the $ \pm 10\% $ uncertainty in the distance.  

\subsection{Fitting results}
\label{ss:ssfit}

Broad-band fitting of a complex spectrum with a single model for the lines and continuum is generally not a very good idea, since the process in highly non-linear and there may be a lot of local maxima of the likelihood function across the parameter space. We therefore first try to get some insight regarding the probable parameters values from the observables considered in Section \ref{s:observ}.

As a first step, we fitted the HEG spectrum by a power law in the 3-6 keV energy range. This resulted in an excellent fit for the binned data ($ \chi^2\approx  195.08 $ for 194 d.o.f., $ \Gamma=1.40\pm0.07 $, $ A=0.018\pm0.002 $) and yielded almost identical parameter values for the unbinned weighted data. As is seen in Fig.\ref{f:observ}, the derived slope is too flat to be produced by the jet model alone. Hence, some additional component (e.g. reflected disc emission) or Comptonisation of jet emission is still required even for almost edge-on disc orientation.
The total flux in this range is $F_{3-6}=4.70 \times 10^{-11}$ erg/cm$ ^2 $/s, which corresponds to a luminosity $L_{3-6}=1.19 (1.70)~\times 10^{35}$ erg/s for $ d_{SS433}=4.6 (5.5) $ kpc. Hence, the total luminosity $ L_X\simeq 7.4 L_{3-6}=8.8(12.6)\times 10^{35}$ erg/s (see Section \ref{ss:norm}). Assuming equal contributions from the red and blue jets, we find $ L_{k}\geq L_{k,min}\sim 10^3 \times L_{X}/2\sim 4(6)\times 10^{38} $ erg/s per jet. 

Next, we notice the data in the 6-9 keV energy range and fit a power-law model with fixed $ \Gamma=1.40$ but with the addition of an exponential high-energy cut-off (in order to account for possible spectrum curvature) and six broad gaussian lines at the positions expected for the blue- and red-shifted Fe XXV K$\alpha$, Fe XXVI Ly$\alpha$ and  Ni XXVII K$\alpha$ lines plus a narrow Fe I $K\alpha $ line at 6.4 keV. A satisfactory fit ($ \chi^2\approx 349$ for 299 d.o.f.) is achieved for $ E_{cut}\approx 15.5 $ keV. The corresponding flux ratio is $F_{6-9}/F_{3-6}\approx 1.0  $. As can be seen from the second top panel of Fig.\ref{f:observ}, this already implies $ \sim 2$ times solar metallicity even for the highest $ T_0 $, which in turn may be estimated from the $ k= $Fe XXVI Ly$\alpha$/Fe XXV K$\alpha$ flux ratio. For this observation, Fe XXVI Ly$\alpha$ of the red jet and Fe XXV K$\alpha$ of the blue one are blended, and only their combined flux can be measured \citep{Lopez2006}. Nonetheless, assuming that the $ k$-ratio is the same for the blue and red jets, we can recover it by solving the quadratic equation $ Ak^2-Bk+C=0$, where $ A,B$ and $C$ are the photon fluxes in the Fe XXV K$\alpha$ line of the red jet, in the blend, and in the Fe XXVI Ly$\alpha$ line the blue jet, respectively. According to our fit, $ B/A\approx 1.64$ and $ C/A\approx 0.36 $ (cf. with $ B/A\approx 1.7$ and $ C/A\approx 0.33 $ found by \cite{Lopez2006}). Solving the equation for k and choosing the $k<1 $ root (see Fig.\ref{f:observ}), one finds $ k\approx 0.27$ (the results of \cite{Lopez2006} imply that $ k= 0.23^{+0.10}_{-0.07}$). This result indicates a relatively low $ T_0\approx 12$ keV, and also implies that the red jet's flux is reduced by a factor of $ 0.7 $. Further, the Ni XXVII K$\alpha  $ line of the red jet is confidently detected at a level $ \sim 0.4$ of the corresponding Fe XXV K$\alpha  $ line, which indicates a nickel-to-iron abundance ratio $ \sim 10 $ times the solar one. Also, the narrow Fe I $K\alpha $ line at 6.4 keV is detected at a level $ 8\pm2 \times10^{-5}$ ph cm$ ^2$/s, confirming the presence of the additional component due to reflection by relatively cold gas even at an almost 'edge-on' precession phase. Since the contribution of this component is not expected to be important for $E<3 $ keV \citep{Medvedev2010}, and bearing in mind the hints for the plausible parameter values found above, we may now proceed to real fitting of the model to data in the 1-3 keV energy range. Also, it is feasible to make use of the MEG data here and perform a simultaneous fit for the HEG and MEG datasets.

\begin{table}
\topcaption{Results of simultaneous fitting of the SS 433-specific model (see Eq.(\ref{eq:model})) to HEG and MEG datasets (unbinned Churazov-weighted and binned to contain at least 25 counts per bin) in the 1-3 keV energy for the two values of distance to SS 433 $ d_{SS433}=4.6 $ kpc and $ d_{SS433}=5.5 $ kpc. The errors cited for the binned data correspond to $ 1\sigma $ uncertainty. The model is marginally degenerate for $ \tau_{e0}\lesssim 0.02$, hence the global fitting constrains only the $ L_{k}\tau_{e0}$ combination in this regime. }
\begin{tabular}{ccccc}\hline\hline
\multicolumn{5}{c}{Simultaneous HEG \& MEG fits in the 1-3 keV energy range  }\\
\hline
$ d_{SS433} $& \multicolumn{2}{c}{4.6 kpc}  & \multicolumn{2}{c}{5.5 kpc}  \\
\hline
& Unbinned & Binned  & Unbinned & Binned  \\
\hline
\smallskip
$ n_{H}~^a$& 1.33 & 1.32 $ \pm $ 0.04 & 1.33 & 1.33 $ \pm $ 0.04\\
$ z_b $ & 0.0103 & 0.0103  & 0.0103 & 0.0103 \\
$ z_r $ & 0.0618 & 0.0616 $ \pm $ 0.0005 & 0.0618 & 0.0616 $ \pm $ 0.0005\\
\smallskip
$ \Sigma(6keV)^b$ & 24.5 & 24.5  & 24.5 & 24.5 \\
$L_{38}~\tau_{e0}~^c $ & 0.63 & 0.64 $ \pm $ 0.09 & 0.98 & 0.93 $ \pm $ 0.13\\
$T_0~^d$ & 14.9 & 16.4 $ \pm $ 3.8 & 14.3 & 16.2$ \pm $ 3.8\\
$ Z $ & 2.25 & 2.4 $ \pm $ 0.5 & 2.12 & 2.38 $ \pm $ 0.5\\
\smallskip
$ Z_{Ni} $ & 19.6 & 21.3 $ \pm $ 4.7 & 18.4 & 20.7 $ \pm $ 4.5\\
$ const ~^e$ & 0.55 & 0.55 $ \pm $ 0.04 & 0.55 & 0.55 $ \pm $ 0.04\\\hline
$ \chi^2/d.o.f. $ &  & {1003 / 776} &  & {1006 / 776}\\
\hline
\end{tabular}
	$ ^a 10^{22}$ cm $ ^{-2} $;
	$ ~^b$ eV;
	$ ~^c L_{38}=L_k/10^{38}$ erg/s;
	$ ~^d $ keV.\\
	$ ~^e $ An artificial suppression factor for the red jet's normalisation.
\label{t:fits}
\end{table}
 
As a preparatory step, we isolated a region around the Si XIV Ly$ \alpha$ line of the blue jet and fitted the model locally in order to find the redshift and width of the line. This gives $ z_{b}=0.0103 \pm 0.0004 $ and $ \Sigma(6~keV)= 24.5  \pm 2 $ eV (all errors throughout the paper correspond to 1$ \sigma $ uncertainties, unless otherwise stated), which is consistent with the findings of \cite{Lopez2006} (see Section \ref{ss:sssmodel} and Section \ref{ss:xspec}). In what follows, we fix the blue jet's redshift and Gaussian smoothing parameter for both jets at these values, while the redshift of the red jet is left free, since it cannot be accurately measured due to blending of the Si XIV Ly$ \alpha$ line of the red jet with the blue jet's Si XIII K$ \alpha$ triplet. 

The results of the fitting for the binned and unbinned weighted data are summarised in Table \ref{t:fits} for $ d_{SS433}=4.6 $ kpc and $ d_{SS433}=5.5 $ kpc. It appears that the data favour a small $ \tau_{e0}\lesssim 0.02 $, for which the model is marginally degenerate, so it is possible to constrain only the combination $ \tau_{e0}L_{k}$. Fitting the binned and unbinned weighted data yields consistent results for both values of $ d_{SS 433} $, and the best-fitting parameters are almost the same for both $ d_{SS 433} $ except for the normalisation-defining combination $ L_{k}\tau_{e0}$. Moreover, the best-fitting parameters are also broadly consistent with our findings based on observables in the harder (3-6 keV and 6-9 keV) energy bands. The best-fitting model provides $ F_{1.5-3}/F_{3-6}\approx 1.$ for the ratio of the \textit{absorption-corrected} fluxes, so this diagnostic also provides a plausible range of the parameter values (see the third top panel of Fig.\ref{f:observ}). Local fitting of the He-like triplets of silicon and sulphur does not allow breaking the degeneracy due to blending with the Ly$ \alpha $ lines for the blue jet and insufficient quality of the data for the red jet. The latter appears $ \sim 2 $ times weaker than the blue one in the soft band, as indicated by the artificial suppression factor of $ constant\approx 0.55$ for the red jet (see Eq.(\ref{eq:model}) and Table \ref{t:fits}).

Although the model provides quite a good fit in the 1-3 keV energy range, it leaves significant residuals at higher energies (see Fig.\ref{f:obs1019}), as has already been anticipated above from the spectral slope in the $ 3-6 $ keV sub-band. There is a clear peak in the data-to-model ratio corresponding to the Fe K$ \alpha $ line at 6.4 keV, so at least some part of this excess comes from the putative reflection component. However, if the parameter $ constant$ is set to unity while all other parameters are kept the same as for the 1-3 keV best-fitting model, such a model almost fully accounts for the deficit of the original model with respect to the data at high energies (see Fig.\ref{f:obs1019}) and leaves residuals only in the Fe K$ \alpha $ line and adjacent continuum. Such a model also slightly over-predicts the fluxes of the red jet's iron lines. This possibly indicates that emission from a cooler jet region is partially blocked while the hottest region is totally unobscured (see also discussion of this issue in \cite{Lopez2006}). Any further interpretation of these results is beyond the scope of the current paper, while the main conclusion we may draw is that the calculated model is indeed capable of describing the SS 433 jets' emission sufficiently well for a reasonable set of parameter values.   
\begin{figure}
\includegraphics[bb=50 140 520 690,width=0.95\columnwidth]{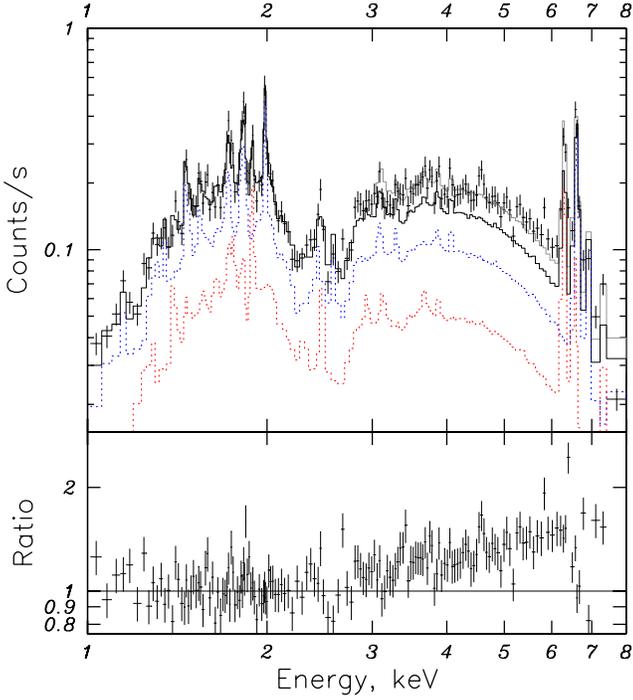}
\caption{The $ Chandra $ HEG data for SS 433 at an almost 'edge-on' precession phase (ObsId 1019, re-binned for better visibility only) with the model providing the best simultaneous fit (for $ d_{SS4333}=5.5$ kpc) to the HEG \& MEG datasets in the 1-3 keV range shown by the black solid line. The blue and red dotted lines show contributions of the blue and red jets respectively. The gray line (drawn only for $ E>3$ keV) shows the same best-fitting model, but with no artificial suppression of the red jet's normalisation (i.e. $ constant$ set to 1).}
\label{f:obs1019}
\end{figure}
\section{Conclusions}
\label{s:conclusions}

We provide the X-ray astronomy community with a spectral model for thermal X-ray emission from a baryonic jet in an XRB, inspired by the unique Galactic microquasar SS 433. The jet is assumed to be strongly collimated (the half-opening angle $ \Theta\sim 1\deg$), mildly-relativistic (the bulk velocity $ \beta\sim 0.03-0.3 $). The X-ray spectrum of the multi-temperature jet is found by summing the contributions from thin slices, each having its own temperature and radiating in the optically thin coronal regime. The temperature profile along the jet and the corresponding differential emission measure (DEM) distribution are calculated self-consistently taking into account cooling due to adiabatic expansion and radiative losses.

The model predicts not only the shape of the emitted spectrum but also the total X-ray luminosity of the jet. Therefore, if the distance to a given source is known sufficiently well, the normalisation of the model is not a free parameter. This makes the overall physical picture self-contained. 

We have also introduced a set of observables that reflect the basic properties of the jet's X-ray emission (such as ratios of fluxes emitted in various energy bands and line to continuum flux ratios) and explored their sensitivity to physical parameters of the jet (such as the temperature and density at its base). Using these observables one can constrain the plausible range of parameter values fairly well without global fitting of a high-resolution X-ray spectrum.  
 
We have checked whether a specific version of our model is capable of fitting high resolution spectra of SS 433 provided by $Chandra$ HETGS for a precession phase when the contribution from non-jet spectral components is expected to be low (when the accretion disc is observed nearly 'edge-on'). The model describes well the data in the 1-3 keV energy range (with the parameter values summarised in Table \ref{t:fits}) but leaves significant residuals at higher energies. The latter can be accounted for if the red jet's emission is modified by a temperature-dependent suppression factor. We also verified the robustness of the fitting results against data representation (binned vs. unbinned weighted) and probed their (in)sensitivity to the assumed distance to the source (except for the normalisation defining combination $ L_{k}\tau_{e0} $). The parameter values inferred from the fitting are broadly consistent with the expectations based on the introduced simple observables.  

Forthcoming data at hard X-ray energies (\textit{NuSTAR}, \cite{Harrison2013}) and with fine spectral resolution near the iron line  complex (\textit{ASTRO-H}, \cite{Takahashi2014}) will be invaluable for checking the consistency of the model's predictions with SS 433 data as well as for assessing its applicability to plausible jet components in the spectra of other Galactic XRBs (e.g. 4U 1630-47, \cite{Diaz2013}), ULXs (e.g. Holmberg II X-1, \cite{Walton2015}), and candidate SS 433-analogues like S26 in NGC7793 \citep{Soria2010} and the unusual radio transient in M82  \citep{Joseph2011}.
   
\section*{Acknowledgements}
~~~~	~~ We are grateful to Eugene Churazov for helpful discussions,
and Max-Planck-Institut f\"ur Astrophysik for hospitality. IK and PM
acknowledge support of Dynasty Foundation.  {The research was
  supported by the Russian Science Foundation (grant 14-12-01315). We
  would like to thank the referee, Herman Marshall, for very careful
  reading and a lot of useful comments and suggestions, which helped
  to improve the paper.} 

CHIANTI is a collaborative project involving George Mason University, the University of Michigan (USA) and the University of Cambridge (UK). 

\section*{Appendix}
\label{s:app}
\subsection*{A1. Model representation and normalisation }
\label{ss:prange}

The model described in the text predicts luminosity $\mathcal{L}_{0}(\mathbb{P})$ and spectral shape $ \varphi_{\mathbb{P}}(E)  $ of the source emission as a function of its intrinsic parameters $  \mathbb{P} $. In practice, however, one typically operates with the corresponding energy (or photon) flux density
\begin{equation}
F(E)=\varphi_{\mathbb{P}}(E)~\frac{\mathcal{L}_{0}(\mathbb{P})}{4\pi d_s^2},
\label{eq:flux}
\end{equation}
where $ d_s $ is the luminosity distance to the source.

This function is to be determined from a so-called table model, defined by a grid of spectra calculated for a range of parameter values. The model prediction for some arbitrary point in the parameter space is found by means of interpolation on a generic grid. This interpolation may be linear or logarithmic with respect to parameter values, but it is always linear with respect to the predicted flux values. Hence, one should avoid large differences between spectra in adjacent points of the parameter grid, otherwise stepwise changes arise in the predicted spectra (like in the case of the \texttt{APEC} model at low temperatures). Fortunately, this can be done by re-normalizing the original model in such a way that the most drastic dependencies on parameters are retained by the normalisation factor.

 Indeed, as shown in Section \ref{ss:param}, $\mathcal{L}_{0}(\mathbb{P})=C(\mathbb{P}) ~\frac{\tau_{e0}L_{k}}{\theta\beta^3}  $, where $C(\mathbb{P})$ depends on $ \mathbb{P} $ only weakly. Hence,   
 \begin{equation}
F(E)=\frac{C(\mathbb{P})}{4\pi}~\varphi_{\mathbb{P}}(E)~\frac{\tau_{e0}L_{k}}{\theta\beta^3~d_s^2}.
\label{eq:fluxnorm}
\end{equation}
Further, in order to avoid manipulation with large numbers, it is convenient to replace $ L_k $ with $ L_{38}=L_{k}/10^{38} ~erg/s $ and $ d_{s} $ with $ d_{10}=d_s/10 ~ kpc$, so we finally get
\begin{equation}
F(E)=1.05\times 10^{-7}~\frac{C(\mathbb{P})}{4\pi}~\varphi_{\mathbb{P}}(E)~\times~\frac{\tau_{e0}L_{38}}{\theta\beta^3~d_{10}^2}=\mathtt{M}(\mathbb{P})\times \mathtt{N}(\mathbb{P},d_s),
\label{eq:fluxnormtech}
\end{equation}
where the $ \mathtt{M}(\mathbb{P})$ function depends on $ \mathbb{P} $ gently but in a very complicated way, while $ \mathtt{N}(\mathbb{P},d_s) $ varies dramatically but in an obvious fashion with model parameters and distance to the source. Thus, one may treat $ \mathtt{M}(\mathbb{P})$ as a 'model',i.e. a function to be tabulated, while $ \mathtt{N}(\mathbb{P},d_s) $ as its (parameter-dependent) normalisation, whose product with the 'model' results in the predicted emission. In \texttt{XSPEC}, this can be handled by loading $ \mathtt{M}(\mathbb{P})$ as an \texttt{atable} model, and determining its \textit{normalisation} parameter from other parameters:  
\begin{equation}
\mathtt{N}(\mathbb{P},d_s)=\frac{\tau_{e0}L_{38}}{\Theta\beta^3~d_{10}^2}, 
\label{eq:norm}
\end{equation}
which is readily realised by the corresponding parameter links.
 Since $ \beta $ and $ \Theta $ are fixed for the SS 433-specific model (see Section \ref{ss:sssmodel}), its normalisation could be re-defined in terms of $ L_{38}$ , $ \tau_{e0} $ and $ d_{10}$ kpc only:
\begin{equation}
\mathtt{N}(\mathbb{P},d_s)=\frac{\tau_{e0}L_{38}}{ d_{10}^2} 
\label{eq:normss}
\end{equation}
while the corresponding $\mathtt{M}(\mathbb{P}) $  is multiplied by a $ \Theta\beta^3$ factor.

\subsection*{A2. Parameter ranges}
\label{ss:prange}

As the model has to be tabulated on a finite grid of parameter values, it is crucial for such a grid to be maximally consistent with the intended use of the model as well as with its intrinsic limitations (see above). SS 433 is the only source with confident detection of baryonic jets, so we ought to rely on extrapolation of the jet properties from SS 433 to the target source population. Primarily, this refers to $ \beta $ and $ \Theta $, for which we allow values from 0.03-0.3 and from 0.003 to 0.03 rad, respectively. It is worth noting that these parameters may in fact be measured directly from Doppler shifts (for $ \beta $ ) and broadening (for $ \Theta $) of emission lines, so they can be frozen at some pre-determined values when fitting data, as in the case of the SS 433-specific model. The metallicity parameter $ Z $ affects mainly intensity of emission lines, and does it in an almost linear manner. So it is allowed to vary in a relatively broad range of values, namely from 0 to 9. The same is true for the $ Z_{Ni} $ parameter of the SS 433-specific model, but it is allowed to vary from 1 to 30. 

As discussed in Section \ref{ss:param}, the range of $ L_k $ from $ 10^{35} $ to $ 10^{41} $  erg/s is fairly sufficient for the current purposes of the model. The range for $ \tau_{e0} $ is adopted from $ 5\times10^{-5} $ to $ 0.5 $, since for small $\tau_{e0} $ the model becomes sensitive mostly to the combination $\tau_{e0}L_{k} $, while for $\tau_{e0} >0.5$ opacity effects may invalidate model predictions (see Section \ref{ss:param}). Besides that, the interpolation accuracy significantly decreases for $\tau_{e0}\sim 1$,  which, along with the limits for $ T_0 $, is considered next.

\subsection*{A3. Interpolation accuracy}
\label{ss:accuracy}

In order to estimate the accuracy loss caused by interpolation, we performed a Monte-Carlo simulation drawing $\thicksim10^6 $ points $\mathbb{P} $ randomly across the parameter space, and for each of them the fractional deviation $\delta(\mathbb{P})=<|F_{i}(\mathbb{P})-F_{0}(\mathbb{P})|/F_{0}(\mathbb{P})>$} was calculated, where $ F_{0}(\mathbb{P})$ is the actual prediction of the model, $ F_{i}(\mathbb{P})$  is the result provided by the interpolation on the grid. Here, $ <..> $ means averaging over the energy range of interest, say from $ E_1$ to $ E_2 $: $<f>=\frac{1}{E_2-E_1}\int_{E_1}^{E_2}f(E)dE $. The distribution function of $ \delta(\mathbb{P})$ depends on the particular choice of the parameters grid, and we iteratively adjust the grid to achieve the highest accuracy for a given number of grid points.

It turned out that the accuracy of interpolation is worst in the (low-$T$, high-$ Z $, high-$ \tau_{e0}$) region of the parameter space, where contribution of line emission to the radiative cooling term and to the predicted spectrum is most prominent. To keep accuracy level at $ \delta\leq 5\% $ without significant increase in the number of grid points, we choose $ T_{0}=7 $ keV and $ \tau_{e0}=0.5 $ as lower (upper) boundaries for $ T_0 $ and $ \tau_{e0} $ respectively, and also make the grid step finest (in the relative sense) close to these boundaries. 
As a result, when the whole energy range was considered, the accuracy of interpolation was better than 8\% for all points, better than 5\% for $ \approx $97\% of points, and the median accuracy is 2\%. For the photon energies $ E<10$ keV, the median accuracy was close to 1\%, while the fraction of points with  $ \delta(\mathbb{P})>5\% $ is vanishingly small.


\end{document}